\documentclass[useAMS,usenatbib]{mn2e}
\usepackage{times}
\usepackage{graphics}
\usepackage{graphicx}
\usepackage{rotate}
\usepackage{epsfig}
\usepackage{lscape}
\usepackage{longtable}
\usepackage{amssymb}
\usepackage{latexsym}  
\usepackage{amsmath}
\usepackage{wrapfig}
\usepackage{natbib}
\usepackage{subfigure}
\usepackage{color}
\usepackage{multicol}

\begin{document}
\pagenumbering{arabic}
%	Introduction
%		1
%		seminar
%			added Loh and Strauss ref
%	Multi-Scale Probability Mapping
%		2, 3
%		seminar
%	Application to SDSS DR7
%		4.2
%	Group and Cluster Catalogue
%		4.3
%		remove reference to appendix
%		remove BFG quotation and details of the biweight estimator
%	Discussion
%		4.4
%	comparisons with other catalogues
%		remove headings for each other catalogue
%\section{Introduction}
%\section{Multi-Scale Probability Mapping}
%\section{Application to SDSS DR7}
%\section{Group and Cluster Catalogue}
%\section{Discussion}
\title[MSPM: groups, clusters and filaments in SDSS]{Multiscale probability mapping: groups, clusters and an algorithmic search for filaments in SDSS}
\author[A. G. Smith et al.]{Anthony G. Smith$^1$\thanks{E-mail: asmith@physics.usyd.edu.au}, Andrew M. Hopkins$^{1, 2}$, Richard W. Hunstead$^1$ and Kevin A. Pimbblet$^{3}$
\\$^1$Sydney Institute for Astronomy (SIfA), School of Physics, The University of Sydney, NSW 2006, Australia
\\$^2$Australian Astronomical Observatory, P.O. Box 296, Epping, NSW 1710, Australia
\\$^3$School of Physics, Monash University, Clayton, VIC 3800, Australia}
\date{Accepted 2011 December 15.  Received 2011 December 15; in original form 2011 August 14
\\The definitive version is available at \textcolor{blue}{www.blackwellsynergy.com} , or will be very soon.}
\label{firstpage}
\maketitle
\begin{abstract}
We have developed a multiscale structure identification algorithm for the detection of overdensities in galaxy data that identifies structures having radii within a user-defined range.
Our ``multiscale probability mapping" technique \textcolor{black}{combines density estimation with a shape statistic to identify local peaks in the density field.
This technique takes advantage of a user-defined range of scale sizes, which are used in constructing a coarse-grained map of the underlying fine-grained galaxy distribution, from which overdense structures are then identified.}
In this study we have compiled a catalogue of groups and clusters at $0.025 < z < 0.24$ based on the Sloan Digital Sky Survey, Data Release 7, quantifying their significance and comparing with other catalogues.
Most measured velocity dispersions for these structures lie between 50 and 400 km s$^{-1}$.
A clear trend of increasing velocity dispersion with radius from 0.2 to 1 $h^{-1}$ Mpc is detected, confirming the lack of a sharp division between groups and clusters.
A method for quantifying elongation is also developed to measure the elongation of group and cluster environments.
By using our group and cluster catalogue as a coarse-grained representation of the galaxy distribution for structure sizes of $\lesssim 1$ $h^{-1}$ Mpc, we identify 53 filaments (from an algorithmically-derived set of 100 candidates) as elongated unions of groups and clusters at $0.025 < z < 0.13$.
These filaments have morphologies that are consistent with previous samples studied.
\end{abstract}
\section{Introduction}
%	background
%	algorithms
%		empirical / using astrophysical data
%			order - RS then BCG then C4 then High level
%		geometric
%			single-scale
%			multi-scale
Galaxy groups and clusters are important in studies of galaxy evolution and cosmology, with large samples necessary to draw robust conclusions about the role played by environment.
Cluster cores are populated by redder galaxies than elsewhere, and contain a higher fraction of ellipticals, with a corresponding deficit in the number of spirals and irregulars (Dressler \citeyear{Dre1980}).
This trend weakens with increasing redshift, implying evolutionary processes (Dressler et al. \citeyear{Dre1997}) that are dependent on the environmental density of galaxies (Smith et al. \citeyear{Smi2005}).
There is also a star formation rate-density (SFD) relation: cluster cores contain redder galaxies with lower star formation rates.
This result appears to be a continuous function of density, rather than a discrete step divided into ``cluster" or ``void" environments (Hashimoto et al. \citeyear{Has1998}).
The SFD relation is both redshift-dependent (Wilman et al. \citeyear{Wil2005}; Poggianti et al. \citeyear{Pog2006}; Elbaz et al. \citeyear{Elb2007}) and scale-dependent (Balogh et al. \citeyear{Bal2004a}; Kauffmann et al. \citeyear{Kau2004}).

The concentration of mass in a galaxy's environment or host halo (Haas, Schaye \& Jeeson-Daniel \citeyear{Haa2012}) may trigger local physical processes that influence evolution.
Such processes include gas stripping (Gunn \& Gott \citeyear{Gun1972}), shocks in the intracluster medium (Moran et al. \citeyear{Mor2005}), harassment (Moore et al. \citeyear{Moo1996}) and galaxy-galaxy interaction (Ostriker \& Tremaine \citeyear{Ost1975}).
The extent of this influence, contrasted against differences in the evolution of galaxies with specific masses, has the potential to discriminate between models of galaxy evolution.
%caused either by the density of the environment, or is a property of massive elliptical galaxies which just happen to found at the centres of clusters.

The internal structure of clusters presents an important test of numerical simulations (e.g. Lewis, Buote \& Stocke \citeyear{Lew2003}).
As mass tracers, clusters can also constrain cosmology through their counts as a function of redshift (Evrard et al. \citeyear{Evr2002}; Majumdar \& Mohr \citeyear{Maj2004}) and highlight features of large-scale structure.
Groups and clusters are not isolated, but are connected and arranged in a non-trivial manner by cosmic superstructures, including filaments, walls and superclusters (e.g. de Lapparent, Geller \& Huchra \citeyear{dLa1986}).
Filaments may be traced by groups and clusters (e.g. Connolly et al. \citeyear{Con1996}), and often occupy the spaces between massive clusters (Pimbblet, Drinkwater \& Hawkrigg \citeyear{Pim2004}; Colberg, Krughoff and Connolly \citeyear{Col2005a}).
Similarly, superclusters have been identified as unions of smaller structures (e.g. Einasto et al. \citeyear{Ein2001}), demonstrating the use of group and cluster catalogues for the identification of large-scale structure.

Efficient and objective algorithms are required to find and quantify groups and clusters in galaxy data.
Many such algorithms have been developed or refined in recent times to explore the wealth of data available through galaxy surveys (e.g. Miller et al. \citeyear{Mil2005}; Koester et al. \citeyear{Koe2007a}; Dong et al. \citeyear{Don2008}; Robotham et al. \citeyear{Rob2011}) such as the Sloan Digital Sky Survey (SDSS; York et al. \citeyear{Yor2000}; Abazajian et al. \citeyear{Aba2009}).
There are some features exhibited by most clusters that can be exploited by such algorithms.
For instance, most rich clusters contain a group of early-type galaxies found at the centre: the red sequence, showing up as an overdensity in colour-magnitude space (Gladders \& Yee \citeyear{Gla2000}; \citeyear{Gla2005}).
%Many clusters are centred on a giant elliptical galaxy, the (BCG).

An empirical relation between Brightest Cluster Galaxy (BCG) magnitude and redshift (e.g. \textcolor{black}{Brough et al. \citeyear{Bro2002};} Loh \& Strauss \citeyear{Loh2006}) can be used to select galaxies that have expected BCG properties in redshift, colour and magnitude, accompanied by a spatial overdensity of galaxies (Bahcall et al. \citeyear{Bah2003}).
The galaxies contained in any given cluster tend to have similar star-formation histories and can usually be expected to group together when plotted on colour-magnitude diagrams.
%This ``colour-clustering" is used by Miller et al. (\citeyear{Mil2005}) use  to .
This allows a search for ``colour-clustering" along with spatial clustering on the sky and in redshift (e.g. Goto et al. \citeyear{Got2002}).
\textcolor{black}{The C4 algorithm (Miller et al. \citeyear{Mil2005}) identifies clusters as overdensities in a seven-dimensional position and colour space, thus minimising projection effects.
Although the size of the physical spatial aperture is fixed, C4 is multiscale in the sense that the use of colours allows the detection of structures with a range of sizes.}
%Gladders \& Yee (\citeyear{Gla2000}) use this feature by testing slices looking in slices on a colour-magnitude diagram, and finding this tight sequence is a sign of a cluster.

The search for morphological, colour-magnitude and clustering properties can be combined using high-level algorithms, including maxBCG (Koester et al. \citeyear{Koe2007a}) and matched filter (Postman et al. \citeyear{Pos1996}; Kawasaki et al. \citeyear{Kaw1998}; Kepner et al. \citeyear{Kep1999}; Gilbank et al. \citeyear{Gil2004}; Dong et al. \citeyear{Don2008}).
While such algorithms efficiently detect structures with the properties they are trained to find, they are necessarily less sensitive to structures with different properties.
%Structures without strong colour-clustering, structures missing a dominant BCG, or filamentary structures may be missed
%So, what if there are clusters that don't have very strong colour-clustering, and what if there are clusters that don't have a ?
%And it isn't just clusters - there are also more exotic overdensities out there, things like filaments, which we don't understand well enough to use these properties of the galaxy population to find.

If observed colour-magnitude and morphological properties are ignored or not available, galaxy positions alone must be used.
A simple approach is to smooth the input galaxy distribution (e.g. Gaussian smoothing: Balogh et al. \citeyear{Bal2004a}\textcolor{black}{; Yoon et al. \citeyear{Yoo2008}}).
This smoothing performed on the input galaxies makes the galaxy distribution easier to interpret visually, but in such single-scale smoothing, a scale must be chosen, and different overdensity catalogues or properties are obtained with each possible choice.
If the chosen scale is too small, no structures are identified; but if it is too large, all structures are joined together and become indistinguishable.
%Furthermore, the choice of an arbitrary function as kernel (e.g. gaussian) imposes its own structure on the distribution.
%Here I have a set of random points, and I've done a smoothing on three different scales.
%I didn't have to use a gaussian, I could have chosen some other shape.
% - this is gaussian smoothing with different widths - one gets different answers on different scales.

\textit{Multiscale} algorithms are needed to interpret the information gathered on different scales and to make a choice about which scales are most important at which locations, removing the need for manual inspection of output for many scales.
%we don't have to look at output for ten, or however many scales we thought
%On the smallest scale, the smoothing isn't really telling us anything, it's just highlighting the positions of the original points.
%On this intermediate scale, some of the points have been joined together into single structures.
%On this largest scale, we're starting to see everything joined together, and it's starting to give us less information.
Such multiscale algorithms have already been implemented.
For example, the Minimal Spanning Tree (MST; Barrow, Bhavsar, \& Sonoda \citeyear{Bar1985}) joins together input galaxies such that the total edge length is minimised.
An MST approach can be used to recognise structures by \textit{separation}, the removal of all edges above a separation length, equivalent to the friends-of-friends approach (FoF: Huchra \& Geller \citeyear{Huc1982}; Bhavsar \& Splinter \citeyear{Bha1996}\textcolor{black}{; Berlind et al. \citeyear{Ber2006}}) \textcolor{black}{in which the separation length is implemented by a combination of projected and line-of-sight linking lengths}.
This is a way of identifying structures on a range of scales, but the linking length is not directly tied to the scale of structure sought.
Instead, the linking length effectively sets a threshold in \textit{density}, similar to structures identified by a density threshold in the Delaunay Tessellation Field Estimator (van de Weygaert \& Schaap \citeyear{vdW2009}).
%In order to use the MST/FoF approach to identify structures, a choice must be made that selects structures with an unknown range of scales, if the densities .
Wavelet approaches (Slezak, Bijaoui \& Mars \citeyear{Sle1990}; Escalera \& MacGillivray \citeyear{Esc1995}; Vikhlinin et al. \citeyear{Vik1998}) require the choice of an analysing wavelet, introducing shape-dependence.
%The delaunay tessellation defines cells that are uniquely prescribed by the input particles, in which densities are found and interpolated along edges of the Delaunay Tessellation to produce surfaces.
%This is some centrally concentrated set of points, this is the delaunay tessellation, and this gives one cells in which to find densities, like here and here.
%One associates these densities with those points, and uses the Delaunay Tessellation to interpolate between cell centres to get this nice surface.
%While this approach involves no arbitrary assumptions, and will identify structures on a range of scales, and potentially with a range of shapes, there isn't a clear way to get it to target or be most sensitive to a particular range of scales we might be most interested in.

Our objective is the ability to detect structures having radii within a user-defined range (e.g. for finding clusters rather than features of large-scale structure), and to do this without over-specific assumptions about the properties of the target structures.
We have developed a multiscale algorithm that may be directly tuned to be most sensitive to any given range of scales.
By \textcolor{black}{limiting the arbitrariness of our assumptions where possible}, we aim for generality similar to that of statistical correlation function approaches (e.g. Balian \& Schaeffer \citeyear{Bal1989}; Infante \citeyear{Inf1994}).
While probability and scale values may be used to describe \textit{statistical} properties of the galaxy distribution, we use these quantities to \textit{map} the galaxy distribution by locating overdensities.
Our multiscale probability mapping (MSPM) approach is demonstrated by the identification of groups and clusters, predominantly structures with projected radii less than 1 $h^{-1}$ Mpc.
These are subsequently used in an algorithmic search for filaments.

We introduce our new approach, detailing the algorithm, \textcolor{black}{and its suitability for producing a coarse-grained map of the galaxy distribution,} in \S\,\ref{mspm_section}.
Our implementation with SDSS data is described \textcolor{black}{and our selection choices summarised} in \S\,\ref{application_section}.
The results are presented in \S\,\ref{catalogue_section} as a large (10443) catalogue of galaxy groups and clusters.
Measured structure properties are discussed and results compared with previous studies in \S\,\ref{value_section}, along with an observed correlation between group radius and velocity dispersion.
By using the group and cluster catalogue as a coarse-grained representation of the galaxy distribution, we present in \S\,\ref{filament_section} a quantitative algorithmic approach to identify filamentary structure, and an initial filament catalogue.
Our results are summarised in \S\,\ref{summary_section}.
Where necessary, we have adopted $H_{0} = 100h$ km s$^{-1}$ Mpc$^{-1}$, $\Omega_{M} = 0.3$ and $\Omega_{\Lambda} = 0.7$, though the exact choice of values does not significantly affect results at $0 < z < 0.25$. %this has been checked: plus or minus 0.05 in both values makes an 11% difference at most
Except where otherwise indicated, all distances are comoving.

\section{Multiscale Probability Mapping (MSPM)}
\label{mspm_section}
Our algorithm, MSPM, is able to locate overdensities in galaxy positional data, where overdensities are defined as regions that are more dense than average, more dense than surrounding locations, or both.
As a multiscale algorithm, MSPM is sensitive to both high-density small-scale features and extended regions of intermediate density. % this line is adapted from our 2008 paper
Unlike many previous multiscale approaches, this sensitivity may be \textit{directly} constrained to lie within a user-defined scale range.
\textcolor{black}{Aside from} a sampled scale range and resolution\textcolor{black}{, additional selection choices in our implementation are listed in Section \ref{choices_section}}.

\begin{figure*}
\centering
\includegraphics[scale=1]{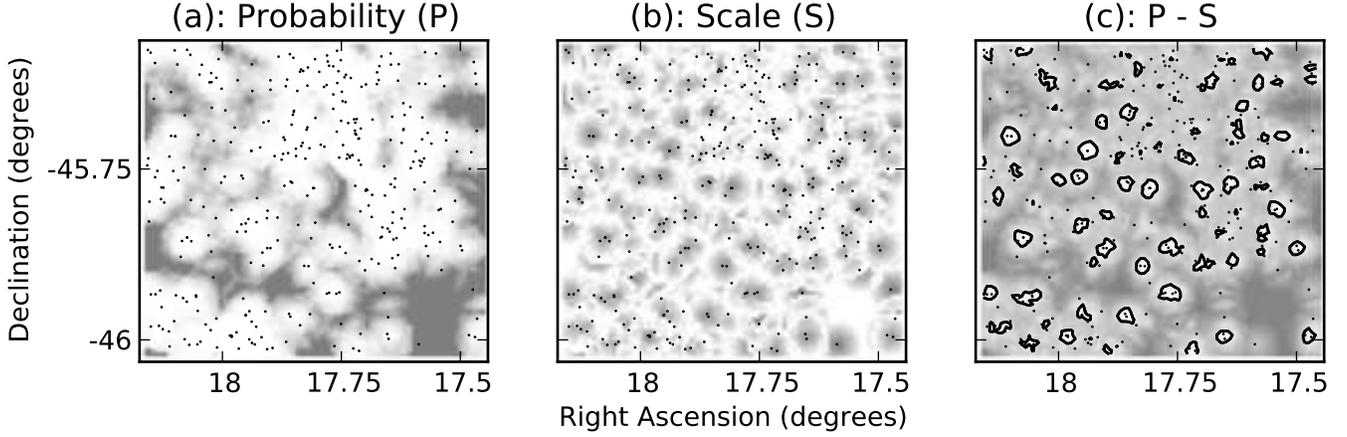}
\caption{A demonstration of MSPM in two dimensions, on a sample of Extremely Red Galaxies (ERGs) studied by Smith et al. (\citeyear{Smi2008}).
The sampled scale range is 20 arcseconds to 2 arcminutes in steps of 10 arcseconds.
Each pixel corresponds to a sampling location on a uniform grid with a spacing of 20 arcseconds.
Black dots are ERG positions.
Sensitivity to isolated galaxies has been reduced.
\textbf{(a)} A probability map showing overdensity probabilities calculated using a Poisson probability density.
High probabilities are displayed as white.
Probability maps are sensitive to overdensities throughout the sampled scale range.
\textbf{(b)} A scale map with high scale values displayed as white.
Scale maps are sensitive to density gradients within high-density islands.
\textbf{(c)} A map showing $P - S$, sensitive to structures that are simultaneously more dense than average and more dense than surrounding locations, subject to the sampled scale range.
Contours enclose structures that would be identified with a threshold $P - S > 0.5$.
Some large structures are missed because they are beyond the sampled scale range.}
\label{probabilityscale}
\end{figure*}

\textcolor{black}{MSPM comprises two distinct parts.
\begin{enumerate}
\item To retain as much useful information about the galaxy distribution as possible while at the same time minimising false detections, a threshold is set in probability rather than density, such that statistically-significant regions are retained, shown in Figure \ref{probabilityscale}(a).
\item To identify structures having radii within a user-defined range, a basic shape statistic is used to identify local peaks in the density field, shown in Figure \ref{probabilityscale}(b).
\end{enumerate}
Together, these two parts provide a method by which to produce a coarse-grained map of the galaxy distribution, that encompasses a range of user-defined scale lengths, shown in Figure \ref{probabilityscale}(c) and discussed in Section \ref{cg_section}.
This output is distinct from smoothing, because the input data are divided into separate regions.
This method of deriving coarse-grained representations can be applied to any distribution of data, not only galaxy positions, and is a technique that MSPM is well-suited to in its implementation.}

A precursor to MSPM has been defined and applied to a sample of Extremely Red Galaxies (ERGs) in the Phoenix Deep Survey by Smith et al. (\citeyear{Smi2008}).
We have enhanced this algorithm by automating structure identification and assignment of scale sizes.
Our complete approach is detailed here.

\subsection{Densities and probabilities}
\label{denprob_section}
The input to MSPM comprises the celestial positions of a set of input galaxies (including redshifts, if available; Section \ref{application_section}), a set of user-defined spatial distances defining a scale range and resolution, and a set of sampling locations.
The scale range should be chosen to extend beyond the largest structures sought if structures are to be selected on the basis of being more dense than their environments (Section \ref{PSthresh}).
The sampling locations may be the galaxy positions themselves (e.g. Kepner et al. \citeyear{Kep1999}), a natural choice that guarantees sufficient and economical sampling of the survey volume.

Our first step is to obtain counts of galaxies around each sampling location.
Each of these counts (densities) $c_i$ is the number of nearby galaxies within a radius equal to one of the user-defined distances $r_i$.
Our use of redshift information in the context of these distances is described in Section \ref{application_section}.
The densities obtained are then converted to probabilities because:
\begin{enumerate}
\item we want to detect structures that are statistically significant,
\item density alone cannot be used to set a threshold throughout a survey that exhibits a varying density (such as the typical variation with redshift resulting from a magnitude-limited survey), and
\item density \textit{contrasts} (e.g. $\rho_\text{inner}/\rho_\text{outer}$) may be undefined in low-density regions (where $\rho_\text{outer} = 0$).
%\item density \textit{contrasts} are a candidate but also vary when density is not fixed
\end{enumerate}

To compute probabilities, counts are compared with probability densities.
Each probability density (e.g. Poisson or Gaussian) is the probability of obtaining each possible count of galaxies.
For meaningful probabilities that take into account the statistics of the galaxy distribution, a natural choice (referred to as ``empirical") is derived from a statistically-comparable ensemble of counts (for example, those obtained at a similar redshift).
Then the probability of obtaining a count $c$ at some individual location, within a specific radius $r_i$, is:
\begin{equation}
\label{indprob}
P_i(c) = \frac{\text{sampling locations with a count $c$ within radius $r_i$}}{\text{number of sampling locations in the ensemble}} .
\end{equation}
With a cumulative probability density $P^\prime_i$ for each radius $r_i$ containing a count $c_i$, we have:
\begin{equation}
\label{intprob}
P^\prime_i = \sum_{c=0}^{c_i-1}[P_i(c)] + \frac{1}{2}P_i(c_i) .
\end{equation}
This is approximately the fraction of the probability density with $c < c_i$, and quantifies the probability of obtaining $c_i$ or less within a radius $r_i$.

$P^\prime_i$ includes $\frac{1}{2}P_i(c_i)$ so that in the case where the $i$-th count is $n$ and all the counts within its ensemble are also $n$, the probability is 0.5\textcolor{black}{, since $P_i(c_i) = P_i(n) = 1$ (equation \ref{indprob})}.
In this way, a probability is associated with each sampling location, creating the $i$-th single-scale probability ``map".
This process is repeated for each of the input scales $r_i$, resulting in a probability map for each scale.
%These counts are compared with what we would expect, given the mean density throughout our area, and from this comparison we compute a probability - the probability of obtaining that count or less, given the average density.
%The higher the density, the higher this probability, so this probability becomes a proxy for density.
%These probabilities are used to compare between scales.

\subsection{Probability and scale maps}
%Finding a probability at each location gives us a smoothed map (e.g. Smith et al. \citeyear{Smi2008}), but unlike gaussian smoothing - we do not assume a shape.
%The generation of probabilities is repeated for each of the input scales (all the scales we wish to be sensitive to), producing a probability map for each scale.
To obtain our final probability map, we assign to each sampling location the maximum value of probability from all of the single-scale measurements at that location:
This gives an \textit{overdensity probability}, $P$, at each sampling location:
\begin{equation}
P = \text{max}(P^\prime_i) .
\end{equation}
If an overdensity is present on any of the sampled scales, it will be evident in this probability map.

The other half of MSPM is \textcolor{black}{the shape statistic, allowing for} scale selection.
The scale, $S$, at each sampling location is defined to be the radius hosting the highest probability ($P^\prime$) at that location:
\begin{equation}
S = r(P^\prime)|_{P^\prime = P} .
\end{equation}
%This slide is the same as the previous one, but here, instead of taking the maximum value from all of these, we remember the maps that gave the highest probabilities, for each location.
%So, these bright patches - at these locations, the map that gave the highest probability was this one, at the top of the scale range.
\textcolor{black}{On this ``scale map", low scale values are usually associated with local density peaks.
For sampling locations close to a peak in the local density field, higher densities and hence higher probabilities $P^\prime_i$ will be obtained at small radii, corresponding to the close proximity of the density peak.
Thus, the highest probability will be found at small $r$, resulting in a low scale value $S$.}

Equivalently, if the single-scale probability values ($P^\prime_i$) at a given sampling location are a probability function (of radius), that function's maximum is the overdensity probability.
The scale value at that location is the radius at which the maximum occurs.
To prevent probabilities rising where the count of objects does not, probability functions are defined to be zero in the absence of additional counts enclosed with increasing radius.

A probability map and a scale map are shown in Figure \ref{probabilityscale}, demonstrating their different but complementary functions.
A probability map by itself may join all structures together since most locations are overdense on at least one scale within a large scale range, or identify structures with boundaries that do not correspond to actual density variations.
The scale map compensates for this behaviour by comparing local densities with surrounding locations, guaranteeing contrasts within the sampled scale range.
More complex distributions than that shown in Figure \ref{probabilityscale} cause the scale map by itself to identify structures that may not be denser than average.
Our work with SDSS does not readily produce images because an adaptive grid is used to reduce computational effort (Section \ref{application_section}).
Hence, the demonstration in Figure \ref{probabilityscale} uses data from the Phoenix Deep Survey (Hopkins et al. \citeyear{Hop2003}; Smith et al. \citeyear{Smi2008}) to create two-dimensional images.

\subsection{Thresholding with $P$ and $S$}
\label{PSthresh}
The next step is to interpret the probability and scale information assigned to each sampling location in the form of a ($P, S$) pair.
Different combinations of the two parameters can be used to locate various features of the galaxy distribution.
%Going over what we have so far - we started off with this original set of points, and from that we found counts at each location and got our probability and scale values, giving us probability and scale maps.

The probability map highlights regions that are overdense when compared to the average density, as measured within the sampled scale range.
Assuming an empirical probability density (Section \ref{denprob_section}), regions above a threshold in probability $P$ contain a densest fraction of the galaxy distribution by number if the sampling locations are the input galaxy positions or by volume if the sampling locations are distributed uniformly throughout the volume.
For example, $P > 0.9$ contains at least the densest 10 per cent in comparison with the mean density.
The fraction selected by a threshold increases as the scale range is increased.
The amount that such a fraction increases also rises as the correlation between large- and small- scale structure decreases.
For instance, $P > 0.9$ selects precisely the densest 10 per cent if only one scale is sampled, and more than 10 per cent as more scales are included in the probability estimator.

The scale map highlights regions that are more dense than surrounding locations over the sampled scale range.
A threshold in scale $S$ will remove sensitivity to an upper portion of the sampled scale range that depends on the threshold.
\textcolor{black}{For example, $S < 0.5$ removes all sampling locations where there is a peak in the probability function $P^\prime_i$ in the upper half of the sampled scale range.
This situation tends to occur when the highest densities are found at large radii, meaning that the sampling location in question is in a relatively underdense region for radii up to the peak in the probability function.}
Assuming an empirical probability density, regions below a threshold in $S$ contain a densest given fraction of the galaxy distribution in a manner similar to a threshold in probability.

$P$ and $S$ do not always correlate.
A structure may, for example, be more dense than surrounding locations but underdense relative to the mean density.
To guarantee the detection of structures that are overdense compared with both the mean density and surrounding locations, both maps are required.
%are actually different things, and we want to use both of these maps.
%We wouldn't want to use either one of them alone - we wouldn't want to use the probability map by itself, because, for example, the probability map is telling us that this, and this, are single structures.
%But we can see that actually there is substructure in there - and the scale map can see this - it shows contrasts within these two regions.
%At the same time, we wouldn't want to use just the scale map, because it doesn't give any credit to regions for being above the average density.
%So for example, in here it isn't picking up on this high density region (remember that high numbers on the scale map mean low densities!), because this is a high density region within a high density region, rather than a high density region within a low density environment.

$P$ and $S$ are interpreted jointly by subtracting $S$ from $P$ (where $S$ is normalised to lie between zero and one).
Subtraction is used because high densities are associated with low values on the scale map ($S$).
Subtraction is preferable to division because it gives the two attributes of being denser than average and more dense than surrounding locations roughly equal weight: identical intervals in $P$ and $S$ usually contain equivalent fractions of the input galaxy distribution.
A $P - S$ map is shown in Figure \ref{probabilityscale}(c).
%This equal weighting results from counts as functions of $P$ and $S$ being constants, for input distributions with small- and large-scale .
%A threshold in ``$P - S$" selects a densest given fraction of the distribution by $P$ and by $S$.
%\footnote{We get all this, and the only input into the algorithm was the initial distribution of points, and the scale range we wanted to be sensitive to.
%The idea is that if one's algorithm is smart enough, that information isn't really necessary.}

While various thresholds in $P$, $S$ or $P - S$ may be motivated by reasoning based on known properties of the target subset of the galaxy distribution, natural choices include:
\begin{enumerate}
\item $P > 0.5$ -- denser than average,
\item $S < 0.5$ -- denser than surrounding locations, and
\item $P - S > 0.5$ -- denser than average \textit{and} denser than surrounding locations, guaranteeing $P > 0.5$ and $S < 0.5$.
\end{enumerate}

%To make this a little clearer, imagine this is a one-dimensional density field.
%This is an overdensity, and this is an overdensity which is a lot more dense.
%If we use just the probability map, we set a probability threshold, which is like setting a threshold in density - imagine this as the horizontal line, our ``sea level".
%Using probability alone, we identify this island as a single structure.
%But it's not just one structure, it's really two; the probability map can't see the substructure.
%The scale map will identify these two banks because they are more dense than surrounding regions.
%But we don't want to just use the scale map because then we will also identify this lower density bank, because even though it isn't really very dense at all, it's more dense than surrounding locations.
%So we want to use both of these maps, and we contour over P - S, probability minus scale.
%
%So far, what I've been showing you have been images which are made from probability and scale maps where we have data everywhere; we have found counts at every location throughout the area along a regular grid, and that makes it easy to do contouring and identify structures.
%But we don't always have that luxury.
%When we're doing this with, for example, with Sloan data, where we might have hundreds of thousands of galaxies, we can't afford to do that, it'd take too long.
%So instead, what we do is only collect data at the places where there are points in the input dataset - we use the input data as an adaptive grid; when we're doing this with real data, we use the galaxies as our adaptive grid.

\subsection{Structure identification}
%This is a contrived set of points meant to reproduce the galaxy two-point correlation function.
%What we do is collect data at all these places, where there are these points.
%At these places we find counts, compute probabilities and scales, and from there we get our P minus S values.
Extended structures are identified as unions of sampling locations (galaxies) above a chosen threshold using a friends-of-friends approach to linking, and are not allowed to contain galaxies below the threshold (e.g. Section \ref{slices+dist}).
The linking length used is the maximum of the sampled scale range.
If the galaxy locations are used as an adaptive grid, the linked locations are identified as member galaxies.
The centre of a structure may be associated with the peak of the region above the threshold, and the distance from this peak to the furthest member galaxy is a measure of radius (Section \ref{catalogue_section}).
%Each of the circular subregions here contain positions that are above the threshold.
%The lines joining these subregions connect things that are close enough together for us to consider them as single structures.
%So that's the tool we're using.

\subsection{Creating coarse-grained distributions}
\label{cg_section}
\textcolor{black}{MSPM is well-suited to constructing a coarse-grained representation of the galaxy distribution in ways that previous algorithms are not.
This is because $S$ is a basic shape statistic, and $P$ allows us to set a low threshold that considers the statistical significance of structures.}

\textcolor{black}{Choosing a limit of $S < 0.5$ (and similarly, $P - S > 0.5$) selects regions that are local peaks in the density field, limiting the ``grain" size of the coarse-grained distribution to the sampled scale range.
This feature is demonstrated in Figure \ref{probabilityscale}(b), in which the grains (dark patches) are not allowed to merge on large scales.
A threshold in density (as approximated by $P > 0.5$; Figure \ref{probabilityscale}(a)) has the potential to join high-density islands together such that the grain-size is not uniform, and not directly controlled by the user.
Since the real galaxy distribution contains structures with a variety of densities, an effective coarse-grained map should be sensitive to shape (as realised by $S < 0.5$), such that densities relative to surrounding locations (as well as to the mean density) are considered.}

\textcolor{black}{Such an approach should attempt to retain as much useful information about the galaxy distribution as possible, while reducing the level of noise.
We would therefore like to set a low threshold while at the same time minimising false detections.
In MSPM, this is attempted by thresholding with $P$ rather than density, such that statistically-significant regions are retained.
Figure \ref{probabilityscale}(a) shows that this has an effect reminiscent of smoothing.
However, the regions identified by $P - S$, shown in Figure \ref{probabilityscale}(c), have a characteristic grain size, a result that is not guaranteed by smoothing.}

\textcolor{black}{The following sections focus on the use of MSPM in identifying galaxy groups and clusters, although the coarse-grained representation of the galaxy distribution produced by MSPM has additional functionality.
In Section \ref{filament_section} we outline previous approaches to, and our use of MSPM in, producing such a representation for the purpose of identifying filaments of galaxies.}

\section{Application to SDSS DR7}
\label{application_section}
\subsection{Survey Volume and Search Apertures}
\label{volap}
We apply our MSPM approach initially to SDSS data (Data Release 7; Abazajian et al. \citeyear{Aba2009}) using only the SDSS main spectroscopic sample, omitting the Luminous Red Galaxy sample (Eisenstein et al. \citeyear{Eis2001}).
To guarantee sufficient sampling while reducing computational effort we use the input galaxy locations as an adaptive grid (e.g. Kepner et al. \citeyear{Kep1999}).
Galaxies less than 2 $h^{-1}$ Mpc from the survey edges are excluded as potential structure centres to reduce edge effects and enable comparison of structure densities with neighbouring volumes (Section \ref{ldc_section}).
Because our approach relies on a comparison with counts obtained at a similar redshift, our comparison volumes at all redshifts must be large enough to allow sufficient statistical strength.
We guarantee this by restricting our attention to $z > 0.025$.

We define our galaxy search volumes as cylinders with variable radii on the sky and a fixed line-of-sight interval in redshift, defined as twice an empirically-obtained \textit{redshift radius}.
To locate groups and clusters rather than features of large-scale structure (e.g. Einasto et al. \citeyear{Ein1984}), our transverse radii are set at 0.2 to 2 $h^{-1}$ Mpc in steps of 0.2 $h^{-1}$ Mpc.
This defines the scale range we sample.
Although we are primarily interested in structures with radii $\lesssim 1$ $h^{-1}$ Mpc, we sample the larger scales so we can select structures that are more dense than their environments.
\textit{Components} of larger structures may be detected by this approach.

Intracluster peculiar velocities stretch structures along the line of sight, preventing us from interpreting redshifts as positions in depth on $\sim$ 1 $h^{-1}$ Mpc scales.
Results obtained from a preliminary analysis have been used as a guide to how deep our cylindrical volumes should be in the line of sight.
We find that a redshift radius of 10 $h^{-1}$ Mpc ($\Delta z \sim 0.004$) is suitable throughout our range of redshift, capturing the spread of peculiar velocities present within most structures.
Thus our cylindrical search volumes have fixed line-of-sight depths in redshift of 20 $h^{-1}$ Mpc.
A smaller redshift radius would probably recover most of the same structures, but their velocity dispersions might be underestimated as a result of the removal of galaxies with large peculiar velocities.

Our larger redshift radius may allow some structures to be identified as unions of physically unassociated galaxies across large distances in the line of sight, and we use sigma-clipping (described in Section \ref{vdisp_section}) to mitigate this effect.

\subsection{Redshift slices and inter-galaxy distances}
\label{slices+dist}
Our probabilities are computed using an empirical probability density (Section \ref{denprob_section}).
The probability associated with a particular count is found by comparison with all other counts within a redshift slice of width $\Delta z = 0.005$ ($\sim 14$ $h^{-1}$ Mpc), centred on that redshift.
This width is chosen to provide a large background comparison volume rather than search for nearby galaxies in the line of sight, and is not related to our redshift radius.
For the area of sky available in Data Release 7, $\Delta z = 0.005$ allows sufficient statistical strength, while retaining sensitivity to decreasing mean density caused by incompleteness at higher redshifts.

Since we use the galaxies as an adaptive grid, a ($P, S$) pair is associated with each galaxy, where $S$ is normalised to lie between 0 and 1.
The physical lengths 0.2 $h^{-1}$ Mpc and 2 $h^{-1}$ Mpc are transformed to 0 and 1 respectively.
To locate structures that are overdense when compared with both the mean density and surrounding locations, we set a threshold of $P - S > 0.5$.
This identifies 177675 of the 619234 galaxies (29 per cent) in the original SDSS sample of galaxies brighter than $r = 17.77$ as lying in overdense environments.

\begin{figure}
\centering
\includegraphics[scale=0.55]{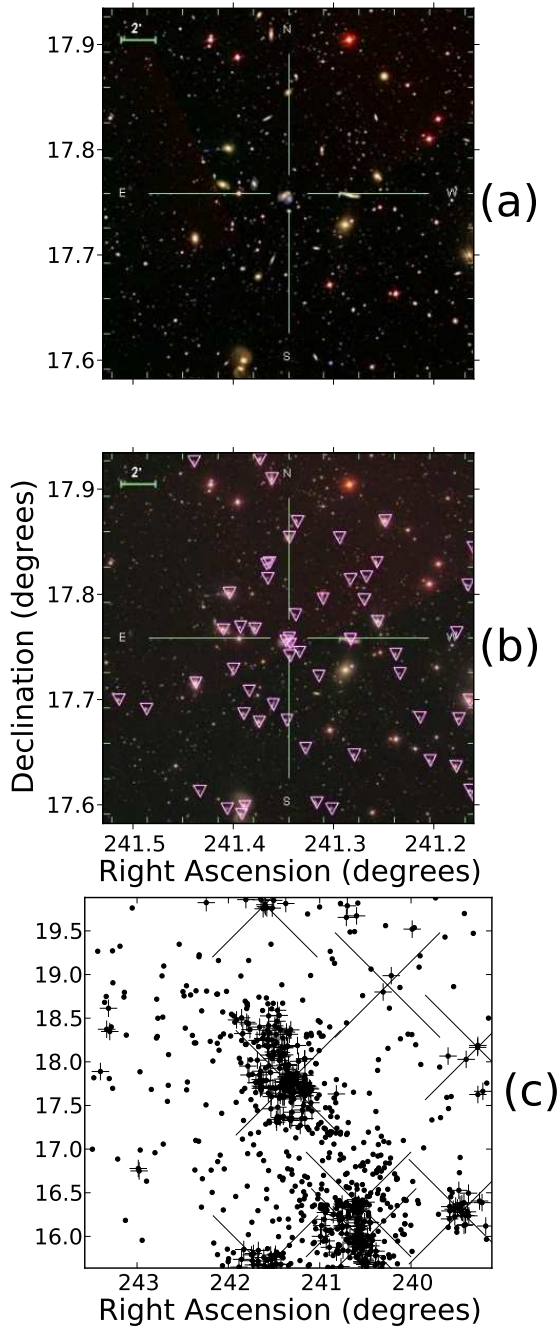}
\caption{The first structure in our catalogue (Table \ref{catalogue}), Abell 2151 (Abell \citeyear{Abe1958}; Corwin \citeyear{Cor1974}) in the Hercules supercluster (Tarenghi et al. \citeyear{Tar1979}).
\textbf{(a)} SDSS image centred on the cluster position, showing a transverse radius of 10.6 arcminutes (0.332 $h^{-1}$ Mpc at $z = 0.036$).
\textbf{(b)} The same image with $r < 17.77$ galaxies within a line-of-sight radius $\Delta z = 0.005$ of the cluster centre marked as triangles.
\textbf{(c)} Objects within a much larger field of view, including a transverse radius of 4 $h^{-1}$ Mpc and the same line-of-sight radius.
Dots are $r < 17.77$ galaxies, plusses are galaxies at positions with $P - S > 0.5$ and large crosses are MSPM structures, some of which are only partially visible within this redshift slice.}
\label{clusterdemo}
\end{figure}

For the purpose of structure identification throughout the survey volume, inter-galaxy distances are defined as
\begin{equation}
\label{metric_eq}
d = \sqrt{d_{\text{t}}^2 + \left(\frac{d_{\text{los}}}{e_{\text{los}}}\right)^2} ,
\end{equation}
where $d_{\text{t}}$ and $d_{\text{los}}$ are the transverse (sky) and line-of-sight comoving separations respectively, and a line-of-sight elongation factor $e_{\text{los}} = 10$ allows a 1 $h^{-1}$ Mpc cluster to contain galaxies apparently up to 10 $h^{-1}$ Mpc away in the line of sight, consistent with our 10 $h^{-1}$ Mpc redshift radius.
Figure \ref{clusterdemo} shows the first structure identified in our catalogue, demonstrating our structure identification on real data.
Galaxies above the threshold are linked together if the distance between them is less than $d =$ 2 $h^{-1}$ Mpc, the maximum of our sampled scale range, and $P - S > 0.5$ for all galaxies (if any) between them, as shown in Figure \ref{clusterdemo}(c).

\subsection{Thresholds}
\label{thresh_section}
We require that each structure, as defined in Section \ref{slices+dist}, has at least four member galaxies with magnitude $r < 17.77$, where each galaxy must have $P - S > 0.5$, so that properties can be measured for each structure.
Collins et al. (\citeyear{Col1995}) find that velocity dispersions determined with fewer than eight radial velocities may be inaccurate, but we have set a lower minimum membership threshold to retain sensitivity to poorer structures.
Our minimum membership requirement excludes 105652 galaxies (out of 177675 with $P - S > 0.5$) that, while individually falling within overdense environments, are not members of such a group.
The minimum overdensity probability of any individual galaxy that is a member of such a group is 0.56.
The density of our structures is compared with their environments in Section \ref{radius_section}.
%This sets an effective minimum overdensity probability of 0.56.
Additionally, we reject structures with low \textcolor{black}{or unmeasurable} local density contrasts and structures with fewer than four member galaxies after line-of-sight sigma-clipping.
These additional criteria are described in Section \ref{catalogue_section}, and affect less than one per cent of our candidate structures.

\subsection{Summary of selection choices}
\label{choices_section}
\textcolor{black}{Although we have tried to minimise the assumptions made about groups and clusters when identifying them, such assumptions are impossible to eliminate entirely.
We have attempted to minimise the arbitrariness of the choices we have made.
Below we list justifications for our more significant choices, along with the selection effects these choices may produce in the resultant catalogue (or cite sections of this paper where they are given).
\begin{enumerate}
\item \textbf{Sampling locations:} by using the input galaxy catalogue as an adaptive grid, we guarantee sufficient sampling while reducing computational effort.
Greater spatial sensitivity would be achieved by use a regular lattice of sampling locations (e.g. Kim et al. \citeyear{Kim2002} for the case of the matched filter algorithm).
The adoption of a regular lattice for MSPM would require changes to our calculation of probabilities to account for the inclusion of void regions.
A suitable adaptation of our approach would recover a similar catalogue of structures.
\item \textbf{Comparison volume:} Local galaxy counts are compared with those obtained from redshift slices of width $\Delta z = 0.005$ (Section \ref{slices+dist}).
From inspecting counts as a function of redshift in the SDSS volume, a narrower slice width would begin to become affected by sample variance (often referred to as cosmic variance) due to sampling an insufficient volume to infer the true local average density.
A larger slice width would introduce biases in the mean density estimate, due to the Malmquist bias resulting from the magnitude limit of the survey.
For example, at the nearer edge of the redshift slice, the average density would be higher than at the farther edge, merely due to the inclusion of intrinsically fainter sources only at the lower redshifts.
\item \textbf{Threshold:} (Section \ref{PSthresh}) $P - S > 0.5$ selects galaxies in regions that are overdense relative to both the mean density and surrounding locations.
29 per cent of our primary galaxy sample satisfies this criterion (12 per cent remain after the minimum count threshold is enforced).
A higher threshold would reduce the false discovery rate in our catalogue but, by including lower-significance detections, the MSPM catalogue retains more information about the galaxy distribution for a study of large-scale structure (Section \ref{filament_section}).
From comparisons with other catalogues, we find that our threshold affects the measured range of velocity dispersions, and hence the masses of the detected structures.
Catalogues constructed from a smaller fraction of the galaxy population contain more massive groups (Section \ref{vdisp_section}).
\item \textbf{Projected scale range:} to locate groups and clusters rather than features of large-scale structure, our transverse sampling radii are set at 0.2 to 2\,$h^{-1}$ Mpc in steps of 0.2\,$h^{-1}$ Mpc.
Our threshold includes regions with $S < 0.5$, corresponding to radii less than 1\,$h^{-1}$ Mpc.
1\,$h^{-1}$ Mpc is the characteristic radius of larger clusters implied by the two-point correlation function (e.g. Einasto et al. \citeyear{Ein1984}).
Sampling larger scales would potentially identify extended structures such as filaments.
\item \textbf{Redshift radius $r_z$:} we find that $r_z =$ 10\,$h^{-1}$ Mpc ($\Delta z \sim 0.004$) is suitable throughout our range of redshift, capturing the spread of peculiar velocities present within most structures, as discussed in Section \ref{volap}.
\item \textbf{Structure identification linking length:} our linking length of 2\,$h^{-1}$ Mpc corresponds to the maximum of our projected scale range.
Since most of our structures have radii $< 1$\,$h^{-1}$ Mpc, altering this large linking length has little effect on the structures we find.
Similarly, our line-of-sight elongation factor $e_{\text{los}} = 10$ allows a 1\,$h^{-1}$ Mpc cluster to contain galaxies apparently up to 10\,$h^{-1}$ Mpc away in the line of sight, consistent with our 10\,$h^{-1}$ Mpc redshift radius.
Altering $e_{\text{los}}$ would thus produce effects similar to those of altering the redshift radius.
\item \textbf{Minimum count of member galaxies:} We require that each structure has at least four member galaxies, as discussed in Section \ref{thresh_section}.
\end{enumerate}}

\section{Group and cluster catalogue}
\label{catalogue_section}
The resultant catalogue in Table \ref{catalogue} contains 10443 structures in the redshift range $0.025 < z < 0.24$, containing a total of 72023 member galaxies, 12 per cent of the input galaxy data.
%sum(C4_nomatch(:,10))/249725=0.0803
This is lower than the 37 per cent identified in the Mr20 catalogue of Berlind et al. (\citeyear{Ber2006}), but greater than the 8 per cent contained by C4 clusters (Miller et al. \citeyear{Mil2005}).
Detailed comparison with these catalogues is discussed in Section \ref{compare_section}.
Structures were sought at $z > 0.24$, but none were found because of incompleteness.
%This is consistent with the classification by Miller et al. (\citeyear{Mil2005}) of 90 per cent of the galaxies in their input data as field-like.

\begin{table*}
\caption{Catalogue of MSPM groups and clusters in SDSS DR7.}
\label{catalogue}
\begin{tabular}{@{}ccccccccccc}
	\hline
                &RA (J2000)&Dec (J2000)&&&&$R$&&&$\sigma_v$&Galaxy\\
                ID&(deg)&(deg)&$z$&$P$&$N$&($h^{-1}$ Mpc)&LDC$_{0.4, 2}$&LDC$_{1, 2}$&(km s$^{-1}$)&$\rho/\bar{\rho}$\\
                (1)&(2)&(3)&(4)&(5)&(6)&(7)&(8)&(9)&(10)&(11)\\
	\hline
1........&241.3437&$17.7596$&0.03629&1.000&157&1.696&$8.6$&$3.9$&680&$513.1$\\
2........&167.7442&$28.6773$&0.03270&1.000&111&0.870&$10.0$&$4.6$&645&$385.0$\\
3........&223.2302&$16.6928$&0.04428&1.000&62&1.555&$8.8$&$5.9$&546&$659.6$\\
4........&240.5182&$15.9474$&0.03341&1.000&47&0.944&$5.0$&$3.6$&408&$532.4$\\
5........&169.1441&$29.2692$&0.04651&1.000&66&0.941&$10.0$&$6.7$&456&$583.6$\\
6........&240.5750&$16.3662$&0.03889&1.000&48&0.903&$4.5$&$3.5$&272&$379.7$\\
7........&234.9231&$21.7713$&0.04120&1.000&71&0.782&$9.0$&$4.2$&541&$475.9$\\
8........&351.1126&$14.6395$&0.04003&0.999&53&0.744&$4.8$&$3.0$&715&$391.8$\\
9........&247.1607&$39.5800$&0.03015&0.999&94&1.187&$3.6$&$2.3$&757&$351.6$\\
10........&247.5227&$40.7662$&0.03047&0.999&85&1.056&$4.8$&$2.3$&580&$353.2$\\
........&........&........&........&........&........&........&........&........&........&........\\
1000........&195.9923&$35.3664$&0.03443&0.865&5&0.720&$3.3$&$1.4$&249&$50.4$\\
2000........&178.4335&$22.3690$&0.06570&0.987&5&0.342&$9.6$&$9.6$&185&$118.2$\\
3000........&171.6480&$3.4761$&0.07444&0.907&5&1.078&$18.0$&$2.2$&267&$76.4$\\
4000........&59.8634&$-6.5318$&0.06150&0.750&4&0.625&$12.0$&$3.8$&101&$52.5$\\
5000........&134.5177&$30.3496$&0.08505&0.972&8&0.691&$5.8$&$2.7$&232&$199.2$\\
6000........&122.9627&$30.2907$&0.07563&0.912&5&0.592&$12.0$&$1.5$&190&$119.9$\\
7000........&159.6631&$23.9223$&0.09476&0.676&4&1.066&$16.0$&$4.5$&117&$73.0$\\
8000........&221.1220&$56.1547$&0.11532&0.886&6&1.484&$14.4$&$1.8$&164&$171.5$\\
9000........&166.1805&$4.2125$&0.14423&0.993&4&0.608&$24.0$&$6.0$&178&$334.0$\\
10000........&155.2763&$30.4010$&0.15498&0.999&6&1.327&$19.2$&$2.4$&340&$697.1$\\
	\hline
\end{tabular}

\medskip
Locations and measured properties of MSPM structures.
Entries are ordered by $P - S$ within slices of ascending redshift, where each slice has a width $\Delta z = 0.025$.
This table shows only a portion of our catalogue as an indication of its content.
The complete catalogue can be found in the online edition of the Journal\textcolor{black}{, or at http://www.physics.usyd.edu.au/sifa/Main/MSPM/ , along with three-dimensional visualisations}.

The selection criteria are described in Section \ref{application_section}.
Columns (2) to (4): position; (5): overdensity probability at peak $P - S$; (6): count of galaxies with $r < 17.77$; (7): radius enclosing region with $P - S > 0.5$; (8) to (9): local density contrasts; (10): velocity dispersion\textcolor{black}{; (11): galaxy density within 0.4 $h^{-1}$ Mpc in units of the background density}.
Our measurements are detailed in Section \ref{catalogue_section}.
\end{table*}

For each structure, we measure a range of properties.
Positions, overdensity probabilities, counts and radii result directly from our structure identification procedure.
Local density contrasts are a density-based quantification of the significance of our detections and velocity dispersions measure the total mass in the systems we have found.

\subsection{Position}
Consistent with our structure selection, we quantify structures without the use of colour-magnitude information, using only our $P - S$ values and identified member galaxies.
An MSPM structure is defined to have the same position on the sky as its member galaxy with the highest $P - S$ value.
For structures that are elongated or asymmetric, an average position on the sky may not select the densest part of the structure.
Peaks in $P - S$ most accurately identify the centres of structures containing at least eight member galaxies.

\begin{figure}
\centering
\includegraphics[scale=0.4]{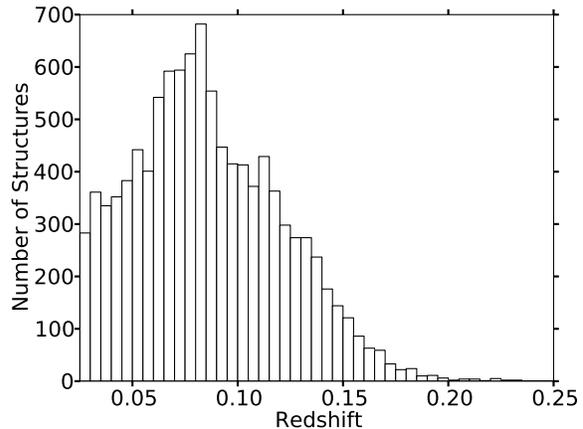}
\caption{The redshift distribution of MSPM structures in Table \ref{catalogue}; mean: 0.086, median: 0.082.
We restrict our attention to $z > 0.025$ (Section \ref{volap}), and do not find any structures at $z > 0.24$ because of incompleteness.}
\label{histz}
\end{figure}

Our approach is analogous to the maximum density measure of Yoon et al. (\citeyear{Yoo2008}).
Since redshifts cannot be interpreted as precise positions on scales of $\sim$ 1 $h^{-1}$ Mpc, our structures have been assigned the average redshift of the member galaxies, similar to Berlind et al. (\citeyear{Ber2006}).
%median(ztemp(find(ztemp>0.025)))       
%    0.1003
%mean(ztemp(find(ztemp>0.025)))  
%    0.1074
Figure \ref{histz} shows that the redshift distribution of our catalogue peaks at $z \sim 0.08$, lower than that for the input galaxy data ($z \sim 0.1$).
The difference is caused by the SDSS magnitude limit.
The consequence of a magnitude limit is for the higher-redshift galaxies that enter the sample to be more luminous and massive, and typically to lie in overdense regions.
However, the fainter members of such overdensities may not enter the sample, and as a result this reduces the number of overdensities that can be recovered at $z > 0.15$.

%plt.ioff()
%plt.figure(num=None)
%plt.hold(True)
%
%centres=(numpy.linspace(0.025+0.0025,0.25-0.0025,45))
%heights=[283,361,335,352,383,442,401,542,592,594,625,682,554,447,415,413,372,429,363,298,274,274,237,176,144,121 ,86 ,63 ,59 ,33 ,22 ,24 ,10 ,11  ,6  ,2  ,4  ,4  ,1  ,5  ,2  ,2  ,0  ,0  ,0]
%width=centres[1]-centres[0]
%plt.bar(centres,heights,width=width,bottom=0,align='center',orientation='vertical',log=False,color='w',edgecolor='k')
%
%plt.axis([0.025,0.25,0,700])
%plt.tick_params(which='both',direction='out',length=5,width=2)
%plt.xlabel('Redshift',size=20)
%plt.ylabel('Number of Structures',size=20)
%
%plt.xticks(size=20)
%plt.yticks(size=20)
%----
%plt.savefig("histz.eps",bbox_inches='tight',pad_inches=0.1,dpi=1,format='eps')

\subsection{Overdensity probability}
\begin{figure}
\centering
\includegraphics[scale=0.4]{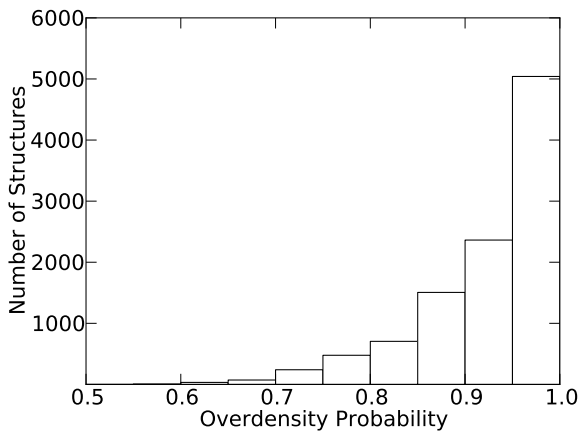}
\caption{Histogram of overdensity probabilities for catalogued structures, median: 0.946.
To qualify for inclusion in our catalogue, each structure must have an overdensity probability of at least 0.5, so that they are in the densest half of the galaxy distribution.
71 per cent of our structures are detected with probabilities greater than 0.9.}
\label{histprob}
\end{figure}

The overdensity probability ($P$) reported for each structure in Table \ref{catalogue} is defined to be the value at its $P - S$ peak.
Figure \ref{histprob} shows that 71 per cent of our structures are detected with overdensity probabilities of 0.9 or greater, meaning that they are in the densest 10 per cent of the galaxy distribution, within the scale range sampled.
Because the centres of overdensities have low $S$ values, most (76 per cent) of these high probability values result from our chosen probability density (Section \ref{denprob_section}) over a radius of 0.2 $h^{-1}$ Mpc, centred at the $P - S$ peak.
%plt.ioff()
%plt.figure(num=None)
%plt.hold(True)
%
%centres=(numpy.linspace(0.525,0.975,10))
%heights=[0    ,6   ,33   ,72  ,240  ,477  ,705 ,1506 ,2363 ,5041]
%width=centres[1]-centres[0]
%plt.bar(centres,heights,width=width,bottom=0,align='center',orientation='vertical',log=False,color='w',edgecolor='k')
%
%plt.axis([0.5,1,numpy.finfo(float).eps,6000])
%plt.tick_params(which='both',length=10)
%plt.xlabel('Overdensity Probability',size=20)
%plt.ylabel('Number of Structures',size=20)
%
%plt.xticks(size=20)
%plt.yticks(size=20)
%----
%plt.savefig("histprob.eps",bbox_inches='tight',pad_inches=0.1,dpi=1,format='eps')

\subsection{Count}
\label{section_count}
\begin{figure}
\centering
\includegraphics[scale=0.4]{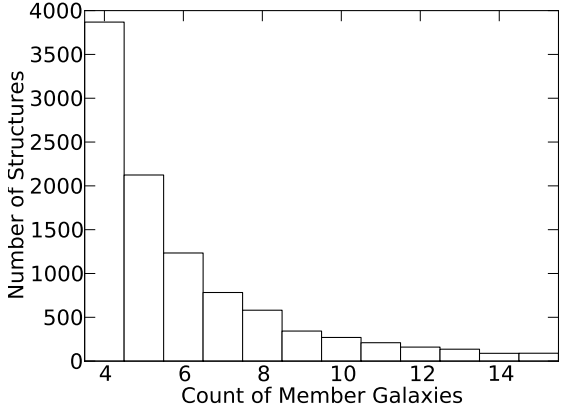}
\caption{Histogram of member galaxy counts for catalogued structures, median: 5.
To qualify for our catalogue, each structure must contain at least four member galaxies (Section \ref{thresh_section}); 37 per cent of our catalogue has this count.
Five per cent of our structures have 16 or more member galaxies (not shown).}
\label{histcount}
\end{figure}

The count of member galaxies is the number of galaxies associated with each structure by our structure identification process.
Each member galaxy must have $P - S > 0.5$.
Figure \ref{histcount} shows that most of our structures have fewer than eight members with $r < 17.77$ above this threshold.
Although we find that counts of member galaxies and overdensity probabilities are correlated, 56 per cent of structures with only four member galaxies still have probabilities greater than 0.9.

%plt.figure(num=None)
%plt.hold(True)
%
%centres=(numpy.linspace(4,20,17))
%heights=[3869 ,2124 ,1234  ,783  ,581  ,343  ,270  ,210  ,160  ,136   ,89   ,90   ,92   ,52   ,43   ,45  ,322]
%width=centres[1]-centres[0]
%plt.bar(centres,heights,width=width,bottom=0,align='center',orientation='vertical',log=False,color='w',edgecolor='k')
%
%plt.axis([3.5,15.5,0,4000])
%plt.tick_params(bottom='off',length=10)
%plt.xlabel('Count of Member Galaxies',size=20)
%plt.ylabel('Number of Structures',size=20)
%
%plt.xticks(size=20)
%plt.yticks(size=20)

75 per cent of structures with eight or more member galaxies (23 per cent of the full sample) have probabilities greater than 0.95.
These form a high-purity subset of our catalogue, with properties measured more accurately, as a higher detected number of the structure members allows a more robust estimate of their radius and velocity dispersion.

\subsection{Radius}
\label{radius_section}
\begin{figure}
\centering
\includegraphics[scale=0.4]{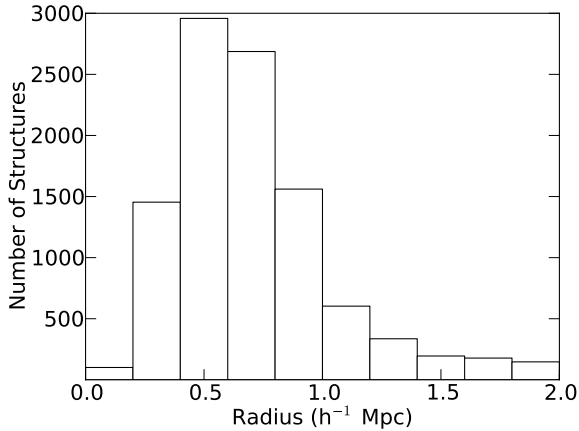}
\caption{Histogram of transverse radii measured for catalogued structures within our sampled scale range; mean: 0.75, median: 0.65.
Two per cent of our structures have radii greater than 2 $h^{-1}$ Mpc (not shown).}
\label{histscale}
\end{figure}

Because of the apparent line-of-sight elongation resulting from galaxy peculiar velocities, we cannot use redshift information to determine the total physical extent of structures on $\sim$ 1 $h^{-1}$ Mpc scales.
We use instead the transverse (sky) distance from the $P - S$ peak to the furthest member galaxy.
\textcolor{black}{This measure will be sensitive to random galaxy displacements, but a more significant limitation for most of our structures is their low count of member galaxies.}
Figure \ref{histscale} shows that most of our measured radii fall within our range of sampled scale values, as expected.
We have found that some of our radii are overestimated as a result of contamination by unassociated nearby galaxies.

%plt.ioff()
%plt.figure(num=None)
%plt.hold(True)
%
%centres=(numpy.linspace(0.1,2.9,15))
%heights=[101 ,1454 ,2958 ,2687 ,1561  ,603  ,336  ,195  ,178  ,147   ,45   ,39   ,38   ,18   ,83]
%width=centres[1]-centres[0]
%plt.bar(centres,heights,width=width,bottom=0,align='center',orientation='vertical',log=False,color='w',edgecolor='k')
%
%plt.axis([0,2,numpy.finfo(float).eps,3000])
%plt.tick_params(top='off',length=10)
%plt.xlabel('Radius (h$^{-1}$ Mpc)',size=20)
%plt.ylabel('Number of Structures',size=20)
%
%plt.xticks(size=20)
%plt.yticks(size=20)
%----
%plt.savefig("histscale.eps",bbox_inches='tight',pad_inches=0.1,dpi=1,format='eps')

\begin{figure}
\centering
\includegraphics[scale=0.45]{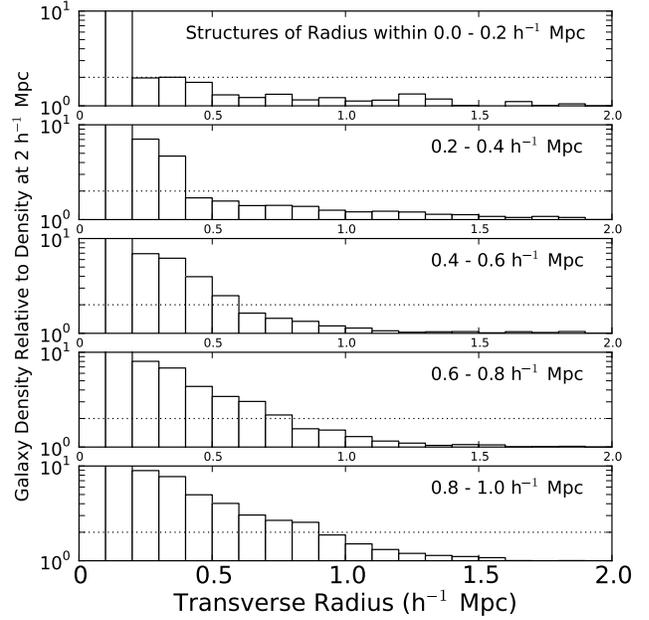}
\caption{Profiles of the galaxy density relative to surrounding locations, including $P - S \leq 0.5$ galaxies, within annuli centred on $P - S$ peaks determined for our structures, averaged over $0.025 < z < 0.2$.
Each panel shows an average of density profiles obtained for all structures with radii within the indicated range.
Horizontal lines indicate twice the density at 2\,$h^{-1}$ Mpc, defined as the density in the annulus formed by rings of radii 1.95 and 2 $h^{-1}$ Mpc.
Relative densities close to group and cluster centres are clipped at 10.}
\label{radiiprofiles}
\end{figure}

To explore the physical significance of the radii we have measured for our structures, we have determined average galaxy density profiles as a function of radius (Figure \ref{radiiprofiles}) for structures with radii up to 1 $h^{-1}$ Mpc.
Galaxies within 10 $h^{-1}$ Mpc in the line of sight are included in the average.
The density profile for each structure is normalised to have a minimum of one, so that all structures have equal weight.
Profiles from individual structures sharing similar radii are then averaged.

Figure \ref{radiiprofiles} shows that our radii approximate boundaries containing regions twice as dense as the density at 2\,$h^{-1}$ Mpc, $\rho_2$, defined as the density in the annulus formed by rings of radii 1.95 and 2.0 $h^{-1}$ Mpc.
Our $P - S > 0.5$ criterion selects regions that are unusually overdense, and this corresponds approximately to regions that satisfy $\rho > 2\rho_2$.
In real space this galaxy density contrast is much higher \textcolor{black}{($\sim 130$, Section \ref{gdc_section})}, without averaging over a large redshift radius.
This is an empirical result, and cannot be generalised to all input distributions.

Our large structures tend to have radii that enclose lower densities relative to the background than those enclosed by the radii of small structures.
Outlying regions of larger structures are above our $P - S$ threshold because they are recognised as extended regions of intermediate density, and thus unusual when compared with most of the galaxy distribution.
The opposite effect occurs for our structures with smaller radii, since large densities at small intergalaxy separations are common, because of the intrinsic clustering of galaxies.
This trend results from our decision to use the locations of the galaxy distribution itself for our probability density measurements.

\subsection{Local density contrast}
\label{ldc_section}
%We define a local density contrast (LDC) as the density within an inner radius of either 0.4 or 1 $h^{-1}$ Mpc projected on the sky divided by the density within the annulus given by an outer radius formed by a 2 $h^{-1}$ Mpc ring.
We define a local density contrast (LDC) as the density within an inner radius divided by the density within the annulus formed by an outer radius projected on the sky.
LDCs are found for two pairs of inner and outer radii: (0.4, 2) $h^{-1}$ Mpc and (1, 2) $h^{-1}$ Mpc.
Galaxies further than 10 $h^{-1}$ Mpc in the line of sight from the structure position are excluded from the galaxy density.
We refer to the LDC measures with subscripts denoting the inner and outer radii in $h^{-1}$ Mpc as LDC$_{0.4, 2}$ and LDC$_{1, 2}$.
LDC measurements allow comparison with the density obtained over a large neighbouring volume.
%We use inner radii of 0.4 and 1 $h^{-1}$ Mpc.
%As mentioned, some counts ($\sim 50$), and hence LDCs, aren't measured because they are incorrectly judged by our method to be close to survey edges; these LDCs are flagged as -1 in our catalogue.

Since small radii centred on individual galaxies are bound to yield high densities, if the count of objects within the inner radius is only one, the resulting LDC is considered unmeasurable and \textcolor{black}{reported in our catalogue as $-1$.
Structures for which neither of our LDCs can be measured are rejected from the catalogue.
Of the structures that remain, where there are no galaxies in the outer annulus, the LDC is undefined, resulting from division by zero.
These structures are retained in the catalogue.}
A larger outer radius or the inclusion of more sensitive observations would resolve this issue.
Our exclusion of all structures close to the survey edges ensures that the outer radius is always within the survey.

\begin{figure*}
\centering
\includegraphics[scale=0.5]{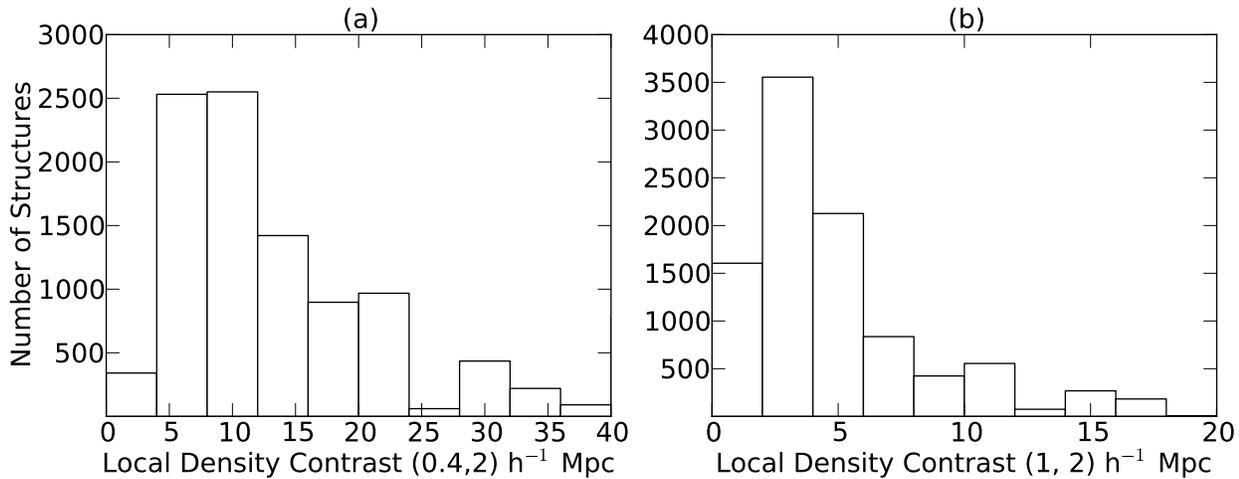}
\caption{\textbf{(a)} Histogram of LDC$_{0.4, 2}$ for catalogued structures; mean: 17.9, median: 12.
Seven per cent of our structures have LDC$_{0.4, 2} > 40$, excluding 107 undefined values not shown (Section \ref{ldc_section}).
\textbf{(b)} Histogram of LDC$_{1, 2}$ for catalogued structures; mean: 5.5, median: 4.
Two per cent of our structures have LDC$_{1, 2} > 20$, excluding 563 undefined values not shown (Section \ref{ldc_section}).}
\label{histldcboth}
\end{figure*}

Defining LDC$_{\text{max}} =$  max(LDC$_{0.4, 2}$, LDC$_{1, 2}$), we require that all of our structures have LDC$_{\text{max}} > 2$.
This constraint excludes 9 structures \textcolor{black}{from our catalogue that are} less than twice as dense as surrounding locations, and all structures that do not have at least two member galaxies within 1\,$h^{-1}$ Mpc of the $P - S$ peak.
On average, LDC$_{\text{max}} = 16.5$, excluding 563 undefined (see above) LDC values.
For 90 per cent of our structures, LDC$_{0.4, 2}$ $>$ LDC$_{1, 2}$.
The jagged appearance of the LDC histograms (Figure \ref{histldcboth}) is a result of integer counts dictating preferred fractions.

\subsection{Global density contrast}
\label{gdc_section}
\textcolor{black}{Densities of our structures in units of the background galaxy density have also been estimated.
The density of a virialised structure in units of the critical density $\rho_\text{cr}$ is predicted by the spherical collapse model (e.g. Bryan \& Norman \citeyear{Bry1998}; King \& Mead \citeyear{Kin2011}) for $\Omega_{M} = 0.3$ and our median redshift $z = 0.082$ to be $\Delta_c = 91$.
This is an overdensity $\Delta = 304$ in units of the background density $\Omega_{M}\rho_\text{cr}$ (e.g. Voit \citeyear{Voi2005}).
Our data allow us to estimate the \textit{galaxy overdensity} rather than the matter overdensity.
Additionally, some galaxies will not be included in the SDSS spectroscopic survey because of fibre collisions, avoidance of bright stars, and other practical survey limitations, so this is not a robust measure of structure densities relative to the mean.}

\textcolor{black}{For each structure, we have calculated a galaxy global density contrast $\rho/\bar{\rho}$ as the density of galaxies within 0.4\,$h^{-1}$ Mpc divided by the density of galaxies within a redshift slice of width $\Delta z = 0.005$ over the entire survey area.
The median global density contrast for our structures is 130.6, and its average is 185.5.
Because of the line-of-sight elongation caused by galaxy radial motions, in counting galaxies within a radius of 0.4\,$h^{-1}$ Mpc we have included galaxies within a line-of-sight radius of 10\,$h^{-1}$ Mpc.
If this count is only one, the resulting global density contrast is considered unmeasurable and reported in our catalogue as $-1$.}

\subsection{Velocity dispersion}
\label{vdisp_section}
Rather than use the radial velocities of all galaxies within an arbitrary transverse radius (such as 1\,$h^{-1}$ Mpc) to calculate velocity dispersions ($\sigma_v$), we use only those galaxies identified by our approach as being structure members.
Equation \ref{metric_eq} allows a large line-of-sight linking length that may contribute member galaxies that are far from structure centres in the line of sight.
These galaxies may not be physically associated with structures, and we use line-of-sight sigma-clipping to remove them, for the purpose of calculating $\sigma_v$ only.
Under this procedure, $\sigma_v$ is calculated using the radial velocities of all member galaxies.
The initial $\sigma_v$ value is used to identify outlying radial velocities, which are then removed before $\sigma_v$ is then recalculated.
This iterative process is also used to remove apparent groups that may result from the chance alignment of unassociated galaxies in the line of sight.

Using the biweight estimator (Beers, Flynn \& Gebhardt \citeyear{Bee1990}), four iterations of $\sigma$-clipping at 2$\sigma$ are applied to the radial velocities.
Our large redshift radius raises the possibility of multiple structures in the line of sight.
To prevent an estimation of $\sigma_v$ for the wrong structure, the median and mean radial velocities are fixed to prevent them from shifting during iteration.
A structure is excluded \textcolor{black}{from our catalogue entirely} if fewer than four member galaxies remain after clipping.
Less than one per cent of our candidate structures are rejected by this criterion.

\begin{figure}
\centering
\includegraphics[scale=0.45]{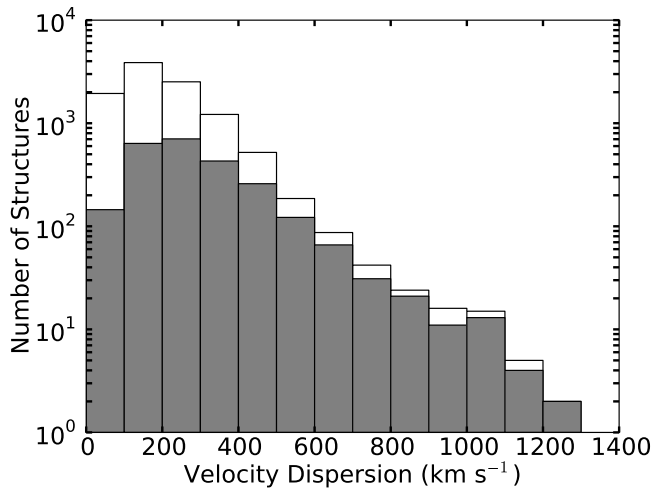}
\caption{\textit{Open histogram:} Velocity dispersions for all catalogued structures; mean: 212 km s$^{-1}$, median: 183 km s$^{-1}$.
\textit{Shaded histogram:} Structures with eight or more member galaxies; mean: 296 km s$^{-1}$, median: 258 km s$^{-1}$.}
\label{histvdisp}
\end{figure}

The range of a catalogue's $\sigma_v$ measurements is strongly dependent on the criteria used to define and select structures.
Our $\sigma_v$ distribution is shown in Figure \ref{histvdisp} and has a median of 183 km\,s$^{-1}$; structures with eight or more members have a median of 258 km\,s$^{-1}$.
The range of our $\sigma_v$ measurements imply masses consistent with structures ranging from poor groups to some of the most massive clusters, with $\sigma_v >$ 1000 km\,s$^{-1}$.
Having allowed relatively poor structures (together containing 12 per cent of the input galaxy data) into our catalogue, we find that our median $\sigma_v$ is comparable with those of samples constructed using similarly low thresholds (e.g. Berlind et al. \citeyear{Ber2006} Mr20: $\sigma_v =$ 128 km\,s$^{-1}$; McConnachie et al. \citeyear{McC2009}: $\sigma_v =$ 227 km\,s$^{-1}$), but lower than those found in studies with higher thresholds (e.g. Miller et al. \citeyear{Mil2005}: $\sigma_v =$ 576 km\,s$^{-1}$).
This is evidence that we are correctly identifying structure members.
Although the range of our results is consistent with previous measurements, most of our individual $\sigma_v$ measurements are based on fewer than eight radial velocities, and may not be accurate.

%plt.ioff()
%
%plt.figure(num=None)
%plt.hold(True)
%
%centres=(numpy.linspace(0.5e5,12.5e5,13))/1000
%heights=[1943,3874,2521,1216 ,520 ,185  ,86  ,41  ,23  ,15  ,14   ,4   ,1]
%width=centres[1]-centres[0]
%plt.bar(centres,heights,width=width,bottom=1,align='center',orientation='vertical',log=True,color='w',edgecolor='k')
%
%heights=[144,636,704,429,258,121 ,65 ,30 ,20 ,10 ,12  ,3  ,1]
%plt.bar(centres,heights,width=width,bottom=1,align='center',orientation='vertical',log=True,color=[0.5,0.5,0.5],edgecolor='k')
%
%plt.tick_params(which='both',direction='in',length=5,width=2)
%plt.xlabel('Velocity Dispersion (km s$^{-1}$)',size=20)
%plt.ylabel('Number of Structures',size=20)
%
%plt.xticks(size=20)
%plt.yticks(size=20)
%----
%plt.savefig("histvdisp_log.eps",bbox_inches='tight',pad_inches=0.1,dpi=1,format='eps')

\section{Purity, radius-velocity dispersion relation and comparisons with other studies}
\label{value_section}
\subsection{Average local density contrast}
\begin{figure}
\centering
\includegraphics[scale=0.4]{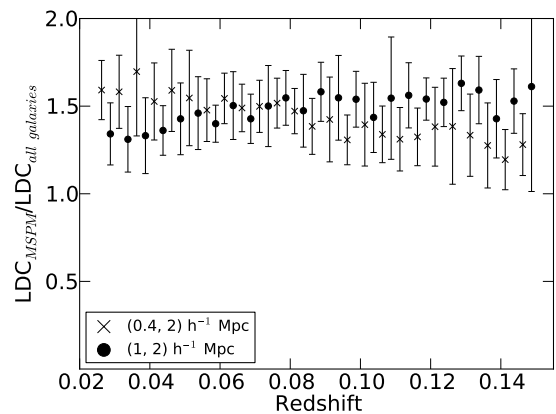}
\caption{LDC values for our catalogued structures divided by the average obtained over all galaxies in the input distribution, for $0.025 < z < 0.15$.
Series are offset for clarity and 1$\sigma$ bootstrap uncertainties are shown.
Note that ``all galaxies" includes those found in our structures, which contain 72023 of the total 619234 galaxies.
\textcolor{black}{Average LDC values for all galaxies are overestimated since they are galaxy-weighted rather than volume-weighted, meaning that dense enviroments are sampled more than poor ones.}}
\label{ldccompareall}
\end{figure}

Because the galaxy correlation function dictates that galaxies \textit{normally} see decreasing densities as a function of radius, high local density contrast (LDC) values are only meaningful if they are also above the average LDC in the general galaxy population.
We compare the LDCs in our catalogue with average LDCs obtained from the whole galaxy population in Figure \ref{ldccompareall}.
At all redshifts, LDCs obtained for our structures are more than those found for the whole galaxy population, within 1$\sigma$ uncertainties.
All these averages are based on LDCs where the inner count of galaxies is at least two.

\subsection{Four-member detections}
%plt.ioff()
%plt.figure(num=None)
%plt.hold(True)
%
%centres=(numpy.linspace(2,42,11))
%heights=[162   ,847   ,786   ,507   ,291   ,398     ,2   ,165    ,90     ,9   ,271]
%width=centres[1]-centres[0]
%plt.bar(centres,heights,width=width,bottom=0,align='center',orientation='vertical',log=False,color='w',edgecolor='k')
%
%plt.axis([0,40,numpy.finfo(float).eps,1000])
%plt.tick_params(which='both',length=10)
%plt.xlabel('Local Density Contrast (max)',size=20)
%plt.ylabel('Number of Structures',size=20)
%
%plt.xticks(size=20)
%plt.yticks(size=20)
%
%plt.savefig("histldcmax_mem4.eps",bbox_inches='tight',pad_inches=0.1,dpi=1,format='eps')
\begin{figure}
\centering
\includegraphics[scale=0.4]{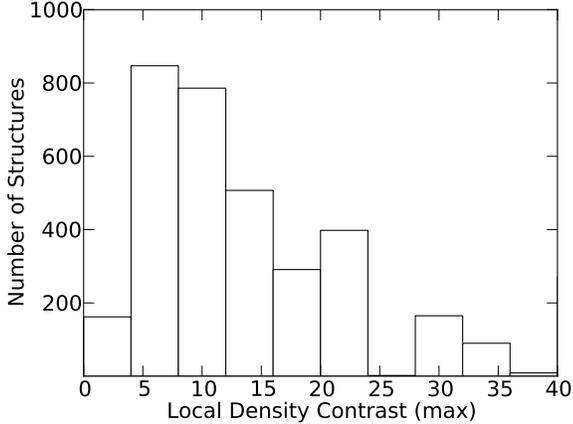}
\caption{LDC$_{\text{max}}$ distribution for structures with only four member galaxies; mean 18.1, median 12.
Six per cent of these four-member detections have LDC$_{\text{max}} > 40$, including 341 undefined values (Section \ref{ldc_section}).}
\label{histldcmax_mem4}
\end{figure}

Over a third of our structures have exactly the minimum count of four member galaxies; the impact of this threshold is discussed in Section \ref{section_count}.
Although this makes them marginal overdensities, they have an average LDC$_{\text{max}}$ value of 18.1, \textit{higher} than that for the remainder of the catalogue, 15.6.
Figure \ref{histldcmax_mem4} shows the distribution of LDC$_{\text{max}}$ values obtained for this subset.
Since four-member structures tend to be surrounded by poorer environments, their high \textit{contrast} values do not indicate higher density than more populous structures.
Higher LDCs may be produced by the relative isolation of a system, its intrinsic richness, or both.

\subsection{Radius-velocity dispersion relation}
\label{vdisprad}
Characteristic properties of groups and clusters are related to their internal dynamics and stages of evolution (Voit \citeyear{Voi2005}).
Older groups are more concentrated, having formed when the universe was denser (Navarro, Frenk \& White \citeyear{Nav1997}), and are more relaxed and spherical than their younger counterparts (Ragone-Figueroa et al. \citeyear{Rag2010}).
Under simplifying assumptions, some properties can be described by scaling relations.
We will treat our groups as isothermal spheres ($\rho(r) \propto r^{-2}$).
More realistic mass-density profiles (e.g. Navarro, Frenk \& White \citeyear{Nav1997}) are shallower at small radii and steeper at large radii.

Assuming an isothermal distribution, velocity dispersion is an indicator of total enclosed virial mass, which should be proportional to $\sigma_v^3$.
This mass is also proportional to the virial radius cubed (e.g. Bryan \& Norman \citeyear{Bry1998}; Kitayama \& Suto \citeyear{Kit1996}), implying a linear relation between radius ($R$) and $\sigma_v$.
If $\Delta_c$ is the mean internal group density in units of the critical density and $H$ is the Hubble parameter, this relation is:
%syms M H Delta G R rho
%vdisp=sym('(M^(1/3))*((H^2*Delta*G^2/16)^(1/6))')
%vdisp=subs(vdisp,M,sym('4*pi*R^3*rho*Delta/3'))
%vdisp=subs(vdisp,rho,sym('3*H^2/(8*pi*G)'))
\begin{equation}
\label{vdisprdelc_eq}
\textcolor{black}{\sigma_v = \frac{1}{2}H\Delta_c^{1/2}R .}
\end{equation}
With the radii we have found for the structures in our sample, we search for a relation between $\sigma_v$ and radius.

\begin{figure*}
\centering
\includegraphics[scale=0.5]{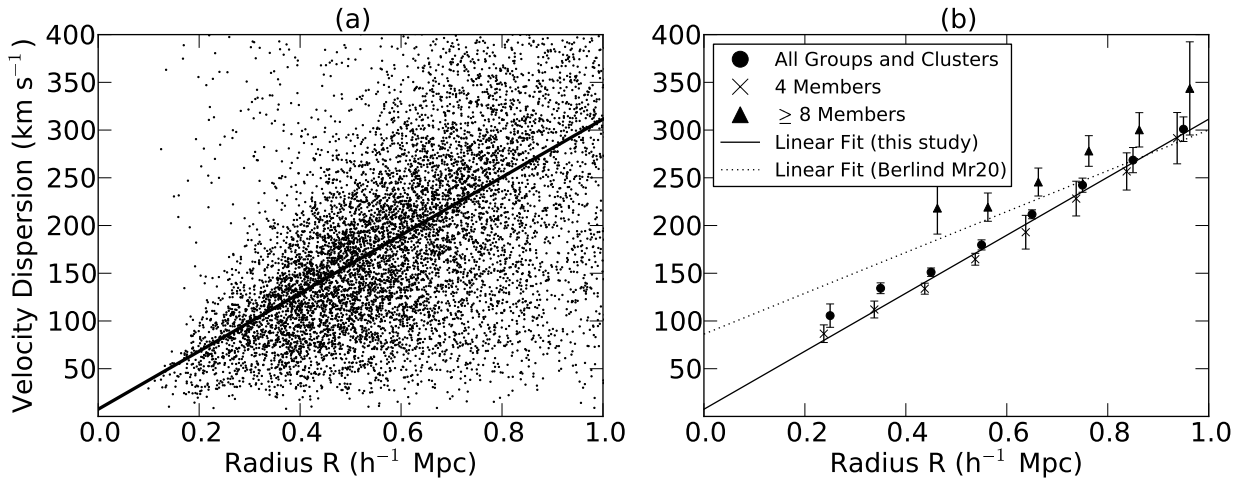} % changed from 0.5 to 0.55 which was too big so changed back
\caption{\textbf{(a)} Velocity dispersion against radius ($R$) for all MSPM groups and clusters\textcolor{black}{, with our linear fit}.
\textbf{(b)} Data points show mean group velocity dispersion against radius with 1$\sigma$ bootstrap uncertainties.
Series are offset for clarity.
Relatively few groups and clusters with at least eight member galaxies have radii less than 0.4\,$h^{-1}$ Mpc, so a reliable average cannot be obtained.
Lines are fits to data with outlying velocity dispersions removed.
See text for details.}
\label{scale_vdisp}
\end{figure*}

Figure \ref{scale_vdisp}(a) shows $\sigma_v$ against radius for all structures in the catalogue (79 per cent of our structures lie within the parameter ranges shown), revealing a clear trend, albeit with much scatter.
The scatter can be partially attributed to uncertainties in the estimates of radius and $\sigma_v$, and to the presence of groups and clusters with various internal densities in our sample (see below).
The velocity dispersions are especially uncertain for low counts of member galaxies.
Radii are determined from the projected separation between the $P - S$ peak and furthest member galaxy, and so are at least uncertain by the mean separation of member galaxies.
Systems in the lower-right corner of Figure \ref{scale_vdisp}(a) have overestimated radii as a result of contamination by nearby unassociated galaxies.

To quantify the trend evident in Figure \ref{scale_vdisp}(a), we have performed a least-squares fit to the data as follows.
%\textcolor{black}{Data for groups and clusters with eight or more member galaxies are removed.
%Figure \ref{scale_vdisp}(b) shows that velocity dispersions for these systems are systematically higher than for less well-resolved systems, increasing the uncertainty in a linear fit.}
Velocity dispersions for all groups and clusters with radii between 0.2\,$h^{-1}$ Mpc and 1\,$h^{-1}$ Mpc are arranged into bins of width 0.1\,$h^{-1}$ Mpc.
Measurements that are more than one standard deviation from the mean $\sigma_v$ in each bin are removed, after which 75 per cent of our data (6537 structures) at $R < 1$ $h^{-1}$ Mpc remain.
Our linear fit with 1$\sigma$ uncertainty, shown in Figure \ref{scale_vdisp}(b), is% this was done with matlab robustfit, polyfit and polyconf
\begin{equation}
\label{vdispfit_eq}
\textcolor{black}{\sigma_v = (304 \pm 3)R + (8 \pm 2)} ,
\end{equation}
where $R$ is the group or cluster radius in units of $h^{-1}$ Mpc and $\sigma_v$ is in units of km s$^{-1}$.
\textcolor{black}{This fit is to the sigma-clipped, unbinned data and is not constrained to pass through the origin.
There may also be underlying systematic uncertainties that are not reflected by the random uncertainties shown in equation \ref{vdispfit_eq}.
The 1$\sigma$ confidence interval for the distribution of data points is $\pm 51$ km s$^{-1}$.}

Although an $R$-$\sigma_v$ relation is not directly noted by Berlind et al. (\citeyear{Ber2006}), we have \textcolor{black}{performed an identical analysis of data from their Mr20 group and cluster sample.
Berlind et al. suggest that their velocity dispersions are 20 per cent too low at all multiplicities, so we apply a 20 per cent upward correction to compensate.
A linear fit through their data at $R < 1$ $h^{-1}$ Mpc has a slope of ($214 \pm 9$) $\text{ }h\text{ Mpc}^{-1}\text{ km s}^{-1}$.
This slope is flatter than ours, but the fit is consistent with our data at $R > 0.5$ $h^{-1}$ Mpc and also shown in Figure \ref{scale_vdisp}(b).}
We stress that close agreement cannot be expected, since $R$ and $\sigma_v$ are calculated differently by Berlind et al., who also set an effectively lower group identification threshold.
This may partially account for their lower value, since the slope of the $R$-$\sigma_v$ relation is a function of the mean group and cluster density in units of the critical density $\Delta_c$ (equation \ref{vdisprdelc_eq}).

\textcolor{black}{To apply equation \ref{vdisprdelc_eq} to our results, a conversion of $R$ to units of Mpc entails division by $h$, removing the need for an assumed value of $H_0$.
We also convert our radii from comoving to proper distances assuming our median redshift $z = 0.082$.
The slope of our $R$-$\sigma_v$ relation and equation \ref{vdisprdelc_eq} implies $\Delta_c = 43.2 \pm 1.0$.
This value is significantly lower than predicted by the spherical collapse model for virialised systems (e.g. Bryan \& Norman \citeyear{Bry1998}; King \& Mead \citeyear{Kin2011}), which for $\Omega_{M} = 0.3$ and our median redshift $z = 0.082$ is $\Delta_c = 91$.}

\textcolor{black}{We note that equation \ref{vdisprdelc_eq} assumes that groups have recently virialised and that $R$ is the virial radius, assumptions we have not examined in this study.
Our low $\Delta_c$ could therefore imply that many of our structures are not virialised or, alternatively, it could arise from systematic underestimation of $\sigma_v$ or overestimation of $R$.
Moreover, the value of $\Delta_c$ implied by the data of Berlind et al. (\citeyear{Ber2006}) is lower than ours, even though Berlind et al. optimised their linking lengths to select group-member galaxies occupying the same virialised dark matter halo.
Hence, the models assumed in determining $\Delta_c$ may be flawed, but we do not examine this issue further.}

A consistent variation of $\sigma_v$ with increasing $R$ to 1\,$h^{-1}$ Mpc is evidence that there is no firm division between groups and clusters.
An $R$-$\sigma_v$ relation could be caused by linking criteria like equation \ref{metric_eq}, but we have ensured that the galaxies included in our velocity dispersion measurements are member galaxies and occupy overdense regions as demonstrated in Figure \ref{radiiprofiles}.
Moreover, sigma-clipping has been used to remove galaxies spuriously included by our large redshift radius and line-of-sight elongation factor.

\subsection{Comparison with other catalogues}
\label{compare_section}
Since we only use data for which spectroscopic information is available, we focus on comparisons with group and cluster catalogues that are similarly derived.
When comparing any two catalogues, appropriate adjustments are made for varying survey areas and varying redshift limits at the time of the catalogue's compilation.
Counterparts are identified by looking within cylinders (aligned with the line of sight) centred at group and cluster centres, with transverse and line-of-sight radii of 1 and 10\,$h^{-1}$ Mpc ($\Delta z \sim 0.004$) respectively.
When determining the fraction of one catalogue recovered by another, the former catalogue's candidates centred less than 2\,$h^{-1}$ Mpc on the sky from the edges of the survey area available to both catalogues are removed so that mismatches are not caused by survey area differences.
Similar adjustments are made to account for the varying redshift ranges of each catalogue.
Our comparisons (with Miller et al. \citeyear{Mil2005}; Berlind et al. \citeyear{Ber2006}; Yoon et al. \citeyear{Yoo2008}) are summarised in Tables \ref{slippers} and \ref{falsepositives}.

%\begin{table}
%\caption{Other catalogues recovered by MSPM.}
%\label{slippers}
%\begin{tabular}{@{}lcc}
%	\hline
%		Other Catalogue&Number&Fraction (\%)\\
%	\hline
%		C4&325/466&70\\
%		Berlind Mr20&1167/1978&59\\
%		Yoon&738/870&85\\
%	\hline
%\end{tabular}
%
%\medskip
%Numbers and fractions of other catalogues that are recovered by MSPM.
%In the case of C4, we compare at $z < 0.1$.
%In the case of Berlind, we consider the subset of Mr20 structures with at least four member galaxies.
%See text for details.
%\end{table}
%\begin{table}
%\caption{MSPM structures found by other catalogues.}
%\label{falsepositives}
%\begin{tabular}{@{}lcc}
%	\hline
%		Other Catalogue&Number&Fraction (\%)\\
%	\hline
%		C4&243/602&40\\
%		Berlind Mr20&1522/2761&55\\
%		Yoon&834/3610&23\\
%	\hline
%\end{tabular}
%
%\medskip
%Numbers and fractions of our MSPM catalogue recovered by other techniques, with the number of MSPM structures adjusted to reflect the %varying survey areas (according to data release) and redshift intervals.
%In the case of C4, we consider the subset of MSPM structures with at least eight member galaxies.
%See text for details.
%\end{table}
\begin{table}
\caption{Other catalogues recovered by MSPM.}
\label{slippers}
\begin{tabular}{@{}lcc}
	\hline
		Other Catalogue*&Number&Fraction (\%)\\
	\hline
		C4&325/466&70\\
		Berlind Mr20&212/394&54\\
		Y08&163/208&78\\
	\hline
\end{tabular}

\medskip
Numbers and fractions of other catalogues that are recovered by MSPM.
In the case of C4, we compare at $z < 0.1$.
In the case of Mr20, we consider the subset with at least four member galaxies.
\textcolor{black}{For Mr20 and Y08 we consider data at $0.09 < z < 0.10$.}
See text for details.
\caption{MSPM structures found by other catalogues.}
\label{falsepositives}
\begin{tabular}{@{}lcc}
	\hline
		Other Catalogue*&Number&Fraction (\%)\\
	\hline
		C4&243/602&40\\
		Berlind Mr20&253/362&70\\
		Y08&162/616&26\\
	\hline
\end{tabular}

\medskip
Numbers and fractions of our MSPM catalogue recovered by other techniques, with the number of MSPM structures adjusted to reflect the varying survey areas (according to data release) and redshift intervals.
In the case of C4, we consider the subset of MSPM structures with at least eight member galaxies.
\textcolor{black}{For Mr20 and Y08 we restrict MSPM to $0.09 < z < 0.10$.}
See text for details.

\vspace{5mm}
*C4: Miller et al. (\citeyear{Mil2005}); Mr20: Berlind et al. (\citeyear{Ber2006}); Y08: Yoon et al. (\citeyear{Yoo2008}).
\end{table}

The C4 catalogue (Miller et al. \citeyear{Mil2005}) is based on DR2 and offers three centroids for sky positions.
In our comparison we consider the peak in the C4 density field since it is the closest analogue to our $P - S$ peaks.
MSPM recovers 62 per cent (431) of the 694 C4 clusters that are more than 2\,$h^{-1}$ Mpc from the survey edges.
Although the C4 catalogue is confined to the spectroscopic data, it uses the LRG sample (Eisenstein et al. \citeyear{Eis2001}), which we have omitted from our input data.
The C4 catalogue is thus based on approximately 1.5 times more data.
At $z < 0.1$, where the LRG fraction has fallen to seven per cent, MSPM recovers 70 per cent (325) of the 466 remaining C4 candidates.

The C4 catalogue imposes a minimum galaxy membership of 8, so to find the fraction of our catalogue matched by C4, we consider the subset of MSPM structures with at least eight member galaxies.
Above the minimum C4 redshift of 0.03, 40 per cent (243) of our structures with eight or more members (numbering 602 in DR2) are matched by C4.
We attribute this low recovery rate to our multiscale approach and to our effectively lower threshold.
Miller et al. (\citeyear{Mil2005}) use apertures with a fixed transverse radius of 1\,$h^{-1}$ Mpc to search for clusters, whereas we search over a range of scales.
Our threshold selects groups and clusters that contain 12 per cent of the input galaxy data, whereas C4 clusters contain 8 per cent.
Our median velocity dispersion for MSPM structures with at least eight members (258\,km s$^{-1}$) is also far lower than in the C4 catalogue (576\,km s$^{-1}$), indicating that we are finding more poor groups.

%REDOING FOR SUBSAMPLES AT 0.09<z<0.1
%how many MSPM does berlind find?
%	restrict mspm (to 0.09<z<0.1), max berlind
%results =
%   1.0e+03 *
%   0253.000000000000   0109.000000000000   0000.698895027624	our catalogue found by the other
%   0328.000000000000   3779.000000000000   0000.079863647431
%how many berlind groups do we find?
%	restrict berlind (to 0.09<z<0.1), max mspm
%results =
%   1.0e+04 *
%   00197.00000000000   10246.00000000000   00000.01886431102
%   00212.00000000000   00182.00000000000   00000.53807106599	other catalogue we find
Berlind et al. (\citeyear{Ber2006}) use a friends-of-friends algorithm to construct a group and cluster catalogue based on DR3 data, using average member galaxy positions for their centroids.
We compare with their Mr20 sample.
\textcolor{black}{Since Mr20 is based on volume-limited input, our comparison is carried out at $0.09 < z < 0.1$ so that our flux-limited catalogue is based on roughly equivalent data.}
MSPM recovers 54 per cent (212) of the 394 Mr20 groups and clusters that are more than 2\,$h^{-1}$ Mpc from the survey edges at $0.09 < z < 0.10$ and that have at least four member galaxies (to match our minimum galaxy membership).
The Mr20 groups that MSPM fails to recover are judged by our approach to be less dense than surrounding locations ($S > 0.5$) within 2\,$h^{-1}$ Mpc.
These groups are still local density peaks when compared with smaller-scale environments.
At $0.09 < z < 0.1$, 70 per cent (253) of the MSPM catalogue (numbering 362 in DR3) is matched by Mr20.

%REDOING FOR SUBSAMPLES AT 0.09<z<0.1
%how many MSPM does yoon find?
%	restrict mspm (to 0.09<z<0.1), max yoon
%results =
%   1.0e+02 *
%   162.0000000000000   454.0000000000000   000.2629870129870	our catalogue found by the other
%   155.0000000000000   769.0000000000000   000.1677489177489
%how many yoon groups do we find?
%	restrict yoon (to 0.09<z<0.1), max mspm
%results =
%   1.0e+04 *
%   00174.00000000000   10269.00000000000   00000.01666187877
%   00163.00000000000   00045.00000000000   00000.78365384615	other catalogue we find
Yoon et al. (\citeyear{Yoo2008}; hereafter Y08) follow a Gaussian weighting scheme to measure densities and construct a cluster catalogue based on DR5 data.
\textcolor{black}{Like the Berlind Mr20 sample, the Y08 catalogue is based on volume-limited input, so our comparison is carried out at $0.09 < z < 0.1$.}
We compare MSPM positions with the sky positions of their maximum-density galaxies and their Gaussian-fitted redshifts, recovering 78 per cent (163) of the 208 Y08 clusters at $0.09 < z < 0.1$ that are more than 2\,$h^{-1}$ Mpc from the survey edges.
%Although the Yoon catalogue allows galaxies in the photometric-only portion of the data to contribute to their detections, this is 
By inspecting mismatches we find that the Y08 clusters we fail to recover are probably real systems, but are not concentrated enough for the MSPM catalogue.
The Y08 catalogue allows galaxies in the photometric-only portion of the SDSS data to contribute to their detections, which may account for at least some of the mismatches.
At $0.09 < z < 0.10$, 26 per cent (162) of the MSPM catalogue (numbering 616 in DR5) is matched by Y08.
\textcolor{black}{A much higher effective threshold is enforced by the Y08 catalogue, which contains approximately three times fewer structures at $0.09 < z < 0.1$ than the MSPM catalogue.}
Moreover, we have sampled a range of scales whereas Y08 follow a Gaussian weighting scheme with a fixed transverse $\sigma$ of 0.7\,$h^{-1}$ Mpc.

Our comparisons show that MSPM recovers most groups and clusters contained in catalogues based on similar data.
However, the relatively low threshold we have set means that many candidate MSPM structures are not detected in other catalogues.
Nevertheless, comparison of our structure LDCs with averages (Figure \ref{ldccompareall}) and the radius-$\sigma_v$ correlation (Section \ref{vdisprad}) are evidence for the reality of the MSPM structures.
Moreover, Figure \ref{radiiprofiles} shows that our groups and clusters are a subset of regions that have twice the density at 2\,$h^{-1}$ Mpc (averaged over a large line-of-sight distance).
It remains probable that false detections and overdensities that are not gravitationally bound have been introduced by our relatively low threshold but, by including lower-significance detections, the MSPM catalogue retains more information about the galaxy distribution for a study of large-scale structure.

\section{Filamentary structures}
\label{filament_section}
We can treat the MSPM group and cluster catalogue (Table \ref{catalogue}) as a coarse-grained representation of the galaxy distribution (Section \ref{cg_section}) with structure sizes of $\lesssim 1 h^{-1}$ Mpc, made possible by our range of sampled scales and threshold in $P - S$.
This demonstrates a use for MSPM's sensitivity to a user-defined scale range.
Treating groups and clusters as particles improves the numerical and computational tractability of large-scale structure studies, suppresses noise contributed by isolated galaxies and reduces the prominence of apparent structures formed by line-of-sight peculiar motions.
Similar approaches have previously been adopted by Colberg (\citeyear{Col2007}) and Zhang et al. (\citeyear{Zha2009}) on simulated data.
Reconstruction of the underlying matter density field from a sample of haloes can also be used to identify and classify features of large-scale structure (Wang et al. \citeyear{Wan2009}; Wang et al. \citeyear{Wan2012}).

Our relatively low threshold for group and cluster identification retains enough information about the galaxy distribution to identify components of filamentary structure.
Our elongation probabilities, introduced below, are a measurement we have devised based on minimal spanning trees to identify filaments as elongated unions of groups and clusters.

\subsection{Identifying and measuring filaments}
Filaments have long been noted (e.g. Kuhn \& Uson \citeyear{Kuh1982}) as prominent features of redshift surveys, and are apparent in deep optical images of fields containing massive clusters (e.g. Kodama et al. \citeyear{Kod2001}; Ebeling, Barrett \& Donovan \citeyear{Ebe2004}).
They have a statistically significant presence (Bhavsar \& Ling \citeyear{Bha1988}) and become prominent when the galaxy distribution is examined on scales above 2 $h^{-1}$ Mpc (Einasto et al. \citeyear{Ein1984}).
However, no entirely algorithmic process has yet been employed to produce a large catalogue of filaments in real data.
A range of algorithms has been suggested for the detection of filamentary structure, including minimal spanning trees (Colberg \citeyear{Col2007}), Delaunay tessellation field estimator (van de Weygaert \& Schaap \citeyear{vdW2009}), modeling as a marked point process (Stoica et al. \citeyear{Sto2005}), skeleton (Novikov, Colombi \& Dor\'e \citeyear{Nov2006}), multiscale morphology filter (Arag\'on-Calvo et al. \citeyear{Ara2007}), ``DisPerSE" (Sousbie, Pichon \& Kawahara \citeyear{Sou2011}) and galaxy axis orientations (Pimbblet \citeyear{Pim2005}).
These algorithms have mostly been applied to simulated data.
Observationally, the difficulty lies in the limitations of real data (completeness, peculiar velocities and projection effects), the lower density of filaments when compared with clusters, and the possibility that simulated filaments are not accurate analogues of actual filaments.
Stoica, Mart\'{\i}nez \& Saar (\citeyear{Sto2010}) suggest that model filaments are shorter than real filaments, and do not form an extended network.

Investigations with both real and simulated data have helped define the properties and morphologies of typical filaments.
In simulated data, filaments are typically 2\,$h^{-1}$ Mpc wide (Arag\'on-Calvo, van de Weygaert \& Jones \citeyear{Ara2010}), tend to have lengths of $\sim 15$\,$h^{-1}$ Mpc (Colberg \citeyear{Col2007}), with a presence that is statistically significant up to a length of $\sim 110$\,$h^{-1}$ Mpc (Pandey et al. \citeyear{Pan2011}).
Arag\'on-Calvo, van de Weygaert \& Jones (\citeyear{Ara2010}) find that more massive clusters host more filamentary connections, and this is supported in real data by Pimbblet, Drinkwater \& Hawkrigg (\citeyear{Pim2004}), hereafter PDH.
Simulated filaments have been morphologically classified by Colberg, Krughoff and Connolly (\citeyear{Col2005a}; hereafter CKC), with results that are also supported in real data by PDH.

We now introduce a metric called the ``elongation probability", which we then use to identify candidate filamentary structures from the MSPM catalogue.

\subsection{Elongation probability}
%doing a convex hull is too hard because every member galaxy will be on the convex hull
%average position to farthest position is not a valid semi-major axis - fitting is the only way down that path
To measure the elongation of a group or cluster's environment, we examine the configuration of nearby groups and clusters by computing their minimal spanning tree (MST\textcolor{black}{; e.g. Barrow, Bhavsar, \& Sonoda \citeyear{Bar1985}).
An MST is a graph that joins together $N$ input particles with $N - 1$ edges such that the total edge length is minimised, without closed circuits.
Overdensities may be identified by the removal of long edges (e.g. Bhavsar \& Splinter \citeyear{Bha1996}), and the distribution of edge lengths may be used to estimate the Hausdorff dimension (e.g. Mart\'{i}nez et al. \citeyear{Mar1990}).
Adjacent edge angles can be used to measure linearity (Krzewina \& Saslaw \citeyear{Krz1996}).
In our approach,} the distribution of angles made by MST edges with a preferred direction is used to calculate an \textit{elongation probability} $P_e$.
%By examining the orientation of MST edges, elongation probabilities are sensitive to the internal alignment of a structure with a preferred direction.

In two dimensions, an MST for an unelongated (isotropic) configuration of structures with $n$ edges can be expected to produce $n/2$ angles less than $\pi/4$ (for example).
Using this \textit{expected count} of angles below the threshold in angle, the \textit{actual count} and a Gaussian probability density, we obtain $P_e$, calculated in the same way as the overdensity probabilities.
For each candidate direction sampled, elongation probabilities are calculated as the average over five angular thresholds linearly spaced between $\pi/20$ and $\pi/4$.

%The preferred direction necessary for the calculation of $P_e$ is the one that maximises it.
%Since this direction cannot be found analytically, it is chosen through trial and error.
Fifteen candidate values of $P_e$ are found under the assumption of each of fifteen directions that are the vectors between six locations:
\begin{enumerate}
\item the average (group and cluster) position,
\item the furthest structure from the average,
\item the structure separated from the average by one quarter of the distance between the average and furthest positions,
\item the structure separated from the average by half the distance between the average and furthest positions,
\item the structure separated from the average by three quarters of the distance between the average and furthest positions, and
\item the position of the structure ranked as having as many structures further away as closer to the average structure position.
\end{enumerate}
\textcolor{black}{This sampling of a limited number of directions reduces computational effort and is analogous to our use of galaxy positions as an adaptive sampling grid when identifying groups and clusters.
In our implementation we are generating MSTs on group and cluster positions rather than individual galaxies, so our MSTs only have six nodes on average.
For these sparse MSTs, the optimal direction will usually align with one of the edges, which will in almost all cases be one of the fifteen directions we sample.
However, a sampling of all directions would allow a better estimate of $P_e$, and should be considered in any future implementation of this approach.}

$P_e$ is calculated for each of these directions, and the maximum value is adopted.
This maximum $P_e$ is always greater than 0.5, and a $P_e$ distribution for MSTs comprising many nodes typically peaks at $\sim 0.75$.
We consider a distribution to be significantly elongated if $P_e > 0.875$.
\textcolor{black}{This is an arbitrary threshold and we have not investigated alternative values.
In the current study it demonstrates the utility of coarse-grained mapping approaches such as MSPM for filament finding.
Any further work should explore the effect of a $P_e$ threshold on purity and completeness, as well as other measures of elongation such as the inertia tensor (e.g. Ragone-Figueroa et al. \citeyear{Rag2010}).}
%the direction sampling has been checked in the code (in all the places it appears) and it is done correctly
% - gaussians aren't quite right for small counts, and even a poisson probability density wouldn't be quite right since the probability of there being more angles above the threshold than there are angles in total is zero.
%But close enough as a nifty add-on tool, I think; as with the comparison of galaxy counts with the ``average", using values from the real data as a probability density is probably the best way.
%Using galaxies with $P - S > 0.5$, the distribution of elongation probabilities is shown in Figure \ref{histelongprob}.

\subsection{MSPM filaments}
\begin{figure*}
\centering
\includegraphics[scale=0.6]{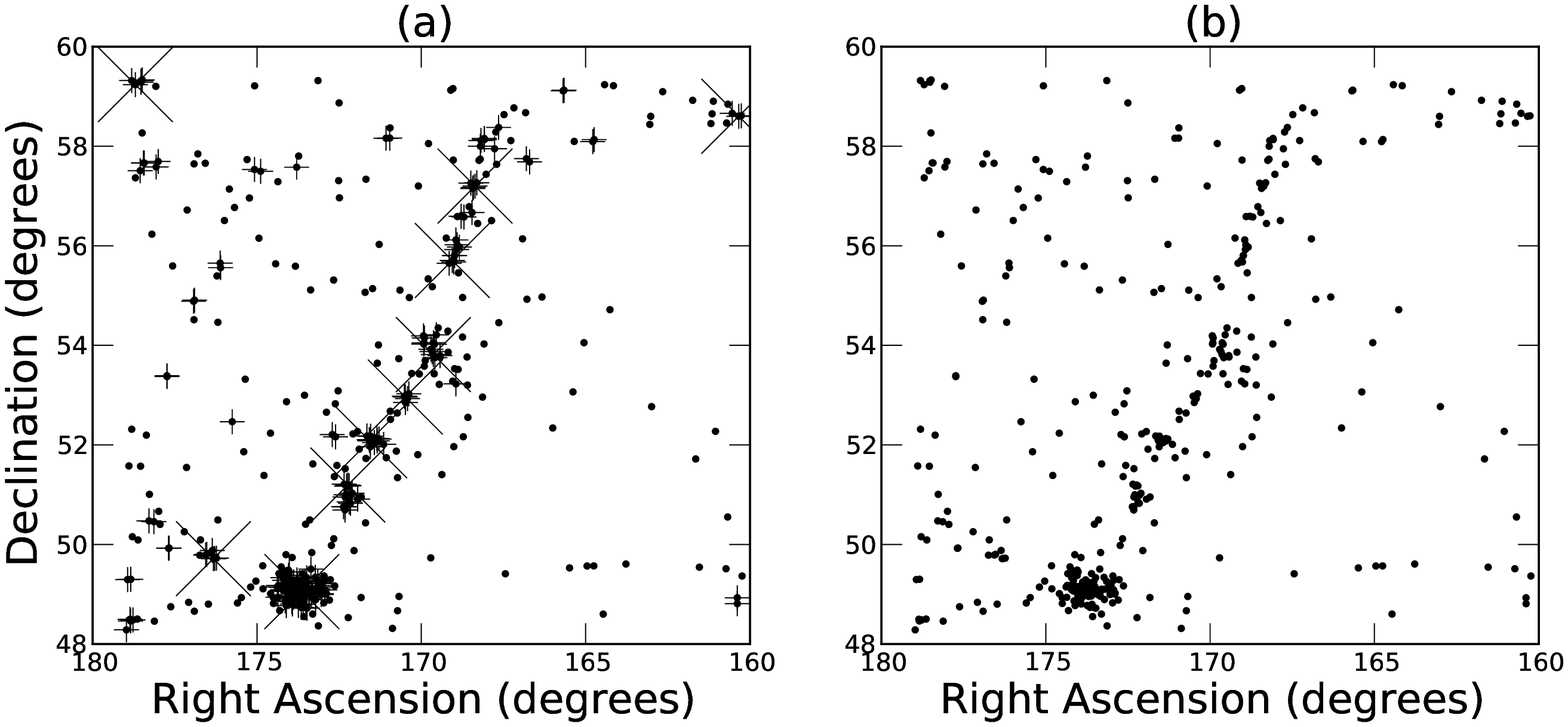}
\caption{A demonstration of our filament detection method, with a filament identified in a field centred on MSPM structure 1063.
Filled circles are $r < 17.77$ galaxies, plusses are galaxies at positions with $P - S > 0.5$ and large crosses are MSPM groups and clusters.
\textbf{(a)} Objects within a transverse radius of 10\,$h^{-1}$ Mpc at $z = 0.03470$ and within a line-of-sight radius $\Delta z = 0.005$.
\textbf{(b)} The same field of view, but with plusses and large crosses omitted.}
\label{filamentdemo}
\end{figure*}

Elongation probabilities are calculated around each of our groups and clusters, using the positions of neighbouring groups and clusters.
An elongation probability is found using structures within each of a set of five radii on the sky: 2 to 10\,$h^{-1}$ Mpc in steps of 2\,$h^{-1}$ Mpc.
Only the sky positions of structures within 10\,$h^{-1}$ Mpc in the line of sight are included, meaning that we are less sensitive to filaments with axes aligned close to the line of sight.
We are not sensitive to thin bridges between groups and clusters less than $\sim 2$\,$h^{-1}$ Mpc long.
An example of a filament traced by MSPM groups and clusters is shown in Figure \ref{filamentdemo}.

Filaments are identified as unions of MSPM groups and clusters that are configured such that their elongation probability is greater than 0.875 for a minimum number $n_{\text{min}}$ of the five sampled scales.
The size (after removal of overlapping volumes) and purity of the resultant filament sample is determined by $n_{\text{min}}$ (Table \ref{mkfilcat}).
Numbers of likely filaments contained by these samples are determined by visual inspection.

\begin{table}
\caption{Size and purity of filament samples.}
\label{mkfilcat}
\begin{tabular}{@{}cccc}
	\hline
		$n_{\text{min}}$&$F_{\text{algorithm}}$&$F_{\text{likely}}$&Purity\\
	\hline
		3&100&53&53\%\\
		4&25&19&76\%\\
	\hline
\end{tabular}

\medskip
Numbers of algorithmically identified ($F_{\text{algorithm}}$) and likely ($F_{\text{likely}}$) filaments contained within, determined by the minimum number of scales $n_{\text{min}}$ with elongation probabilities greater than 0.875.
\end{table}

If four of the elongation probabilities are required to exceed 0.875, an algorithmically-selected sample of filaments is created with a purity of 76\%, but with a sample size of only 25.
For our catalogue and filament morphology work, we have used the 53 apparent filaments identified by visual inspection from the $n_{\text{min}} = 3$ sample (which includes the $n_{\text{min}} = 4$ sample).
The 47 fields that remain were judged to be chance alignments of groups and clusters that do not appear on closer visual inspection to be subunits of filamentary structure.

There are many more filaments manifest in the data that we do not detect, so our filament catalogue's completeness is poor.
Many filaments may fail our $P > 0.5$ requirement since they are not as dense as clusters.
In other cases, the presence of neighbouring structures within 10\,$h^{-1}$ Mpc may lower filament elongation probabilities below our threshold.

Our catalogue of 53 filaments is presented in Table \ref{filcat}.
Each filament is identified by the ID number of the MSPM group or cluster that lies at the centre of the field hosting the filament.
No filaments are detected at $z > 0.13$ because of incompleteness.
%length=2*radius_max(P_e)?
%As a measure of the large-scale environment occupied by filaments, local density contrasts have been measured on large scales.
%Unlike our earlier LDC measurement (Section \ref{ldc_section}), we count \textit{groups and clusters} within inner and outer \textit{spheres} with radii of 10 $h^{-1}$ Mpc and 20 $h^{-1}$ Mpc respectively.
%If the inner count is only one, or if the outer radius is outside the survey edges, LDC$_{10, 20}$ is considered unmeasurable and flagged as $-1$.

\begin{table*}
\caption{Catalogue of MSPM filaments in SDSS DR7.}
\label{filcat}
\begin{tabular}{@{}ccccccc}
	\hline
		&RA (J2000)&Dec (J2000)&&&&\\
		ID&(deg)&(deg)&$z$&$P_e$(max)&$N$&Morphology\\
		(1)&(2)&(3)&(4)&(5)&(6)&(7)\\
% types are by kap. agreed for all but two with ags
	\hline
68........&222.1034&$18.3560$&0.03999&0.924&10&V\\
84........&123.1876&$16.8883$&0.04459&0.904&10&II\\
299........&235.6901&$8.2411$&0.04039&0.894&11&I\\
404........&165.3420&$9.3080$&0.03641&0.908&10&I\\
663........&250.8133&$24.0732$&0.04696&0.905&10&II\\
1063........&169.6346&$53.8145$&0.03470&0.947&7&I\\
1082........&238.3466&$18.3778$&0.03280&0.910&15&V\\
1088........&149.5960&$8.9077$&0.04900&0.921&5&I\\
1494........&196.6982&$60.3546$&0.02784&0.895&6&I\\
1511........&203.9377&$27.8676$&0.02641&0.890&8&V\\
1541........&174.8082&$35.9602$&0.03973&0.904&8&V\\
1946........&208.3502&$19.3225$&0.07120&0.926&6&I\\
2088........&201.0014&$59.0449$&0.07286&0.918&8&I\\
2547........&119.4981&$40.0377$&0.06630&0.908&7&I\\
2770........&174.2258&$44.2294$&0.05886&0.894&4&II\\
2937........&234.3922&$14.3926$&0.05201&0.910&7&I\\
3086........&184.3658&$-0.7805$&0.07020&0.878&3&V\\
3094........&162.2009&$4.3068$&0.06952&0.889&9&V\\
3305........&241.0991&$11.0421$&0.06452&0.924&5&I\\
3502........&223.0171&$17.3097$&0.05819&0.920&9&II\\
3656........&193.7737&$38.6280$&0.05186&0.894&5&II\\
3750........&233.4721&$6.8433$&0.06580&0.878&3&II\\
3861........&195.1090&$52.6712$&0.05473&0.919&8&II\\
4074........&139.3883&$53.2514$&0.05782&0.924&5&II\\
4604........&155.2910&$9.8616$&0.09877&0.898&5&I\\
4647........&175.4488&$56.7553$&0.09737&0.878&3&V\\
4885........&148.4584&$19.9723$&0.08844&0.904&4&I\\
4976........&194.8470&$29.9823$&0.08425&0.878&3&I\\
5021........&175.1702&$10.0262$&0.08225&0.887&9&I\\
5149........&210.5606&$5.9266$&0.07834&0.878&6&V\\
5381........&211.1576&$41.8530$&0.09360&0.878&3&II\\
5527........&183.7894&$36.0110$&0.08906&0.878&3&I\\
5745........&236.1185&$29.6604$&0.08242&0.904&8&V\\
5832........&228.2023&$20.8819$&0.07962&0.920&8&II\\
5853........&226.7288&$7.1890$&0.07958&0.943&13&V\\
5935........&181.9290&$23.8841$&0.07748&0.919&9&I\\
5972........&189.9455&$13.8891$&0.07587&0.905&4&II\\
5984........&183.5435&$17.7961$&0.07697&0.905&4&I\\
6007........&119.5306&$40.8289$&0.07555&0.878&3&I\\
6252........&198.6709&$19.9712$&0.09046&0.919&5&II\\
6280........&229.2817&$3.5373$&0.08093&0.905&7&I\\
6574........&149.8899&$3.1282$&0.08179&0.936&6&II\\
6656........&203.1080&$40.0689$&0.08085&0.878&3&I\\
6695........&135.0231&$53.7212$&0.09165&0.894&4&I\\
6760........&148.4709&$20.6304$&0.07863&0.937&6&II\\
6846........&227.7999&$5.6538$&0.08455&0.905&4&I\\
6858........&170.3387&$38.3315$&0.08609&0.905&6&II\\
6920........&150.8208&$18.5711$&0.07888&0.924&5&I\\
7039........&237.9363&$45.5654$&0.11889&0.878&3&I\\
7642........&139.4112&$36.5945$&0.11032&0.878&3&V\\
8048........&118.0812&$36.1567$&0.11426&0.878&3&II\\
8555........&156.8494&$11.1645$&0.11731&0.921&5&I\\
9625........&163.9845&$40.7248$&0.12916&0.905&4&I\\
	\hline
\end{tabular}

\medskip
Locations, properties and morphologies of MSPM filaments.
Figures showing each filament can be found in the online edition of the Journal\textcolor{black}{, and three-dimensional visualisations can be found at http://www.physics.usyd.edu.au/sifa/Main/MSPM/ .}

Column (1): ID of central MSPM group or cluster; (2) to (4): position; (5): highest elongation probability within 10\,$h^{-1}$ Mpc; (6): count of groups and clusters within a 10\,$h^{-1}$ Mpc radius on the sky and 10\,$h^{-1}$ Mpc in the line of sight (cylindrical aperture); (7): morphological type (Section \ref{filmorphs_section}).
\end{table*}

\subsection{Filament morphologies}
\label{filmorphs_section}
\begin{figure*}
\centering
\includegraphics[scale=1]{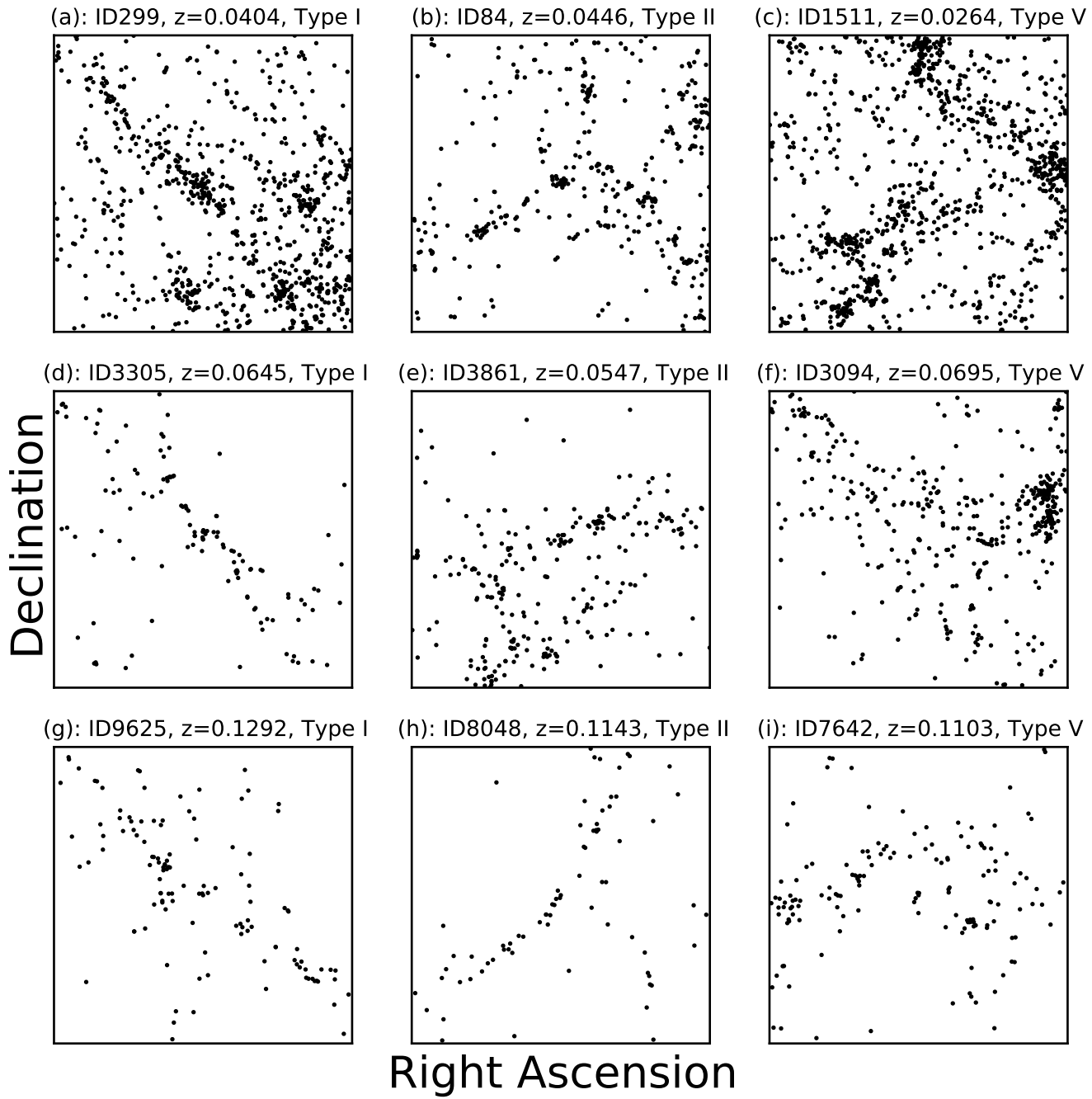}%changed from 0.9 to 1
\caption{Selected fields containing filaments, centred on MSPM groups and clusters.
Each panel shows $r < 17.77$ galaxies within a 20\,$h^{-1}$ Mpc $\times$ 20\,$h^{-1}$ Mpc square on the sky and within a line-of-sight radius $\Delta z = 0.005$ ($\approx 14$\,$h^{-1}$ Mpc).
\textbf{(a)-(c)} Examples of types I, II and V at $z < 0.05$.
\textbf{(d)-(f)} Types I, II and V at $z \gtrsim 0.05$.
Field 3861 shows an example of ``S-shaped" curvature.
\textbf{(g)-(i)} Types I, II and V at $z > 0.1$.
Figures showing all 53 filaments can be found in the online edition of the Journal\textcolor{black}{, and three-dimensional visualisations can be found at http://www.physics.usyd.edu.au/sifa/Main/MSPM/ .}}
\label{showfilaments}
\end{figure*}

Our subjective morphological classification is based on the scheme introduced by PDH and CKC.
For each filamentary field, projections of galaxy positions onto two orthogonal planes are inspected.
In our work, one of these planes is always the sky, since we have only used sky positions in the calculation of elongation probabilities, favouring filaments that are oriented perpendicular to the line of sight.
The other plane is perpendicular to the sky and parallel with the length of the filament.
Our inspections are limited to fields with transverse and line-of-sight radii of 10\,$h^{-1}$ Mpc that may not always capture the endpoints of the filament.
A selection of filaments with different morphologies is shown in Figure \ref{showfilaments}.

In either plane, the projected configuration of galaxies is classified as straight, curved, uniform or irregular.
\begin{enumerate}
\item \textbf{Straight:} galaxies form either a line or lines that are not curved.
\item \textbf{Curved:} a line that is \textit{continuously} bent (not simply crooked) into either a ``C"- or ``S"- shape.
\item \textbf{Uniform:} uniformly distributed galaxies that do not form a clear line.
\item \textbf{Irregular:} irregular distribution of galaxies containing large density fluctuations that obscure the linear structure of the filament.
\end{enumerate}

Following PDH, each filament is assigned a morphological type based on its appearance in two orthogonal planes.
\begin{enumerate}
\item \textbf{Type I (straight):} both are straight (e.g., Figures \ref{showfilaments}a, d, g).
\item \textbf{Type II (warped):} at least one is curved, with neither being uniform or irregular (e.g., Figures \ref{showfilaments}b, e, h).
\item \textbf{Type III (sheet):} one (and only one) is uniform, with the other being either straight or curved.
\item \textbf{Type IV (uniform):} both are uniform.
\item \textbf{Type V (irregular):} both are irregular (e.g., Figures \ref{showfilaments}c, f, i).
\end{enumerate}

Some fields contain multiple filaments, and in these fields we classify the filament containing the groups or clusters that cause a high elongation probability.
Fields are inspected independently by two authors (AGS and KAP) with a 4 per cent disagreement rate.

The division of our filament sample by morphology is shown in Table \ref{filmorphs}, and a selection of filaments with different morphologies is shown in Figure \ref{showfilaments}.
Our results regarding the relative abundance of filament types are consistent with PDH and CKC within uncertainties, finding that most of our filaments are Type I (straight) or II (curved), with the remainder classified as Type V (irregular).
The PDH sample is derived by visual inspection, so our technique is sensitive to the prominent filament morphologies apparent in real data.

\begin{table}
\caption{Filament numbers and fractions by morphology.}
\label{filmorphs}
\begin{tabular}{@{}cccc}
	\hline
		Type&This Study (number)&This Study (per cent)&PDH (per cent)\\
	\hline
		I&26&49 $\pm$ 10&37 $\pm$ 3\\
		II&16&30 $\pm$ 8&34 $\pm$ 3\\
		III&0&0&4 $\pm$ 1\\
		IV&0&0&0.8 $\pm$ 0.5\\
		V&11&21 $\pm$ 6&26 $\pm$ 3\\
	\hline
\end{tabular}

\medskip
The abundance of each filament type in our study compared with PDH.
All uncertainties are Poissonian.
\end{table}

Our Type I fraction of 49 per cent is marginally higher than that found by PDH (37 per cent), who assess volumes that allow greater filament curvature.
PDH record filaments up to a length of $\approx$ 45 $h^{-1}$ Mpc, and find that Type I is the dominant morphology for short ($<$ 10 $h^{-1}$ Mpc) filaments.
Since our search is restricted to fields with radii of 10 $h^{-1}$ Mpc, we do not detect filaments (or filament segments) longer than 20 $h^{-1}$ Mpc.
%The higher relative abundance of straight filaments in our sample reflects the difference in the lengths of our filaments compared with PDH.
Our elongation probability approach is also more efficient at detecting straight filaments than filaments with more complex morphologies.
%The main difference between our work and PDH is that our filaments are not required to extend between clusters.
%Although our filaments comprise groups and clusters, many do not visibly terminate in massive nodes, though some may do so beyond the limits of the fields we have examined.

\section{Summary}
\label{summary_section}
We have designed and implemented a new algorithm, multiscale probability mapping, for the detection of structures in the galaxy distribution.
MSPM can be made sensitive to any chosen range of scales and identifies member galaxies.
Our work with SDSS DR7 data demonstrates its abilities:
\begin{enumerate}
\item by finding groups and clusters with a range of sizes we have quantified a radius-velocity dispersion trend not highlighted in previous work,
\item by identifying groups and clusters through their statistical significance we are able to set a relatively low threshold,
\item MSPM's sensitivity to a user-defined scale range allows us to produce a coarse-grained representation of the galaxy distribution with a user-defined grain size, and
\item using our group and cluster catalogue, we have demonstrated a technique to identify filaments algorithmically with a false discovery rate of less than 50 per cent.
\end{enumerate}

Our filament catalogue omits many filaments present in the data, and we lack an objective way to quantify their morphology.
Future approaches will address these challenges.

The data products made available in this work are a catalogue of 10443 groups and clusters at $0.025 < z < 0.24$ and a catalogue of 53 filaments.
The morphological similarity of our filaments to those of PDH shows that algorithmic filament searches have the potential to produce results comparable to visual inspections of real data.
%Our elongated structure catalogue is the first systematic search for filamentary structure of its kind $<$needs confirmation$>$, and we have started the development of more direct methods in this direction.

\vspace{10mm}
We thank Chris Miller for a useful discussion of cluster finding\textcolor{black}{, and the anonymous referee for detailed comments that helped to improve this paper}.
This research has made use of NASA's Astrophysics Data System, a guide to cosmographic distance measures (Hogg \citeyear{Hog1999}), the VizieR database of astronomical catalogues (Ochsenbein, Bauer \& Marcout \citeyear{Och2000}) and the SIMBAD database, operated at CDS, Strasbourg, France. % used vizier to query catalogues and find a match in Abell; used SIMBAD to look up previous mentions of that cluster
\textcolor{black}{Data products, three-dimensional visualisations and further information about MSPM can be found at http://www.physics.usyd.edu.au/sifa/Main/MSPM/ .}

Funding for the SDSS and SDSS-II has been provided by the Alfred P. Sloan Foundation, the Participating Institutions, the National Science Foundation, the U.S. Department of Energy, the National Aeronautics and Space Administration, the Japanese Monbukagakusho, the Max Planck Society, and the Higher Education Funding Council for England.
The SDSS Web Site is http://www.sdss.org/.

The SDSS is managed by the Astrophysical Research Consortium for the Participating Institutions.
The Participating Institutions are the American Museum of Natural History, Astrophysical Institute Potsdam, University of Basel, University of Cambridge, Case Western Reserve University, University of Chicago, Drexel University, Fermilab, the Institute for Advanced Study, the Japan Participation Group, Johns Hopkins University, the Joint Institute for Nuclear Astrophysics, the Kavli Institute for Particle Astrophysics and Cosmology, the Korean Scientist Group, the Chinese Academy of Sciences (LAMOST), Los Alamos National Laboratory, the Max-Planck-Institute for Astronomy (MPIA), the Max-Planck-Institute for Astrophysics (MPA), New Mexico State University, Ohio State University, University of Pittsburgh, University of Portsmouth, Princeton University, the United States Naval Observatory, and the University of Washington.

\bibliographystyle{scemnras}
\bibliography{X}
\label{lastpage}
\end{document}